\documentclass[12pt]{article}
\usepackage{graphicx}
\usepackage{float}
\usepackage{bm}
\usepackage{caption}
\usepackage{color}
\usepackage{amsmath}
\usepackage[T1]{fontenc} 

\newcommand{\be}{\begin{equation}}
\newcommand{\ee}{\end{equation}}
\newcommand{\ba}{\begin{eqnarray}}
\newcommand{\ea}{\end{eqnarray}}
\newcommand{\n}[1]{\label{#1}}
\newcommand{\eq}[1]{Eq.(\ref{#1})}

\begin{document}

\title{Double Images from a Single Black Hole \!\!
\thanks{Alberta-Thy-16-15}}

\author{Shohreh Abdolrahimi${}^{a}$ \!\!
\thanks{Internet address: abdolrah@ualberta.ca}
 \,  ,
Robert B. Mann${}^{b}$ \!\! 
\thanks{Internet address:
rbmann@uwaterloo.ca}
\, and 
Christos Tzounis${}^{a}$ \!\!
\thanks{Internet address:
tzounis@ualberta.ca}
\\
${}^{a}$Theoretical Physics Institute, \\ 
University of Alberta, Edmonton, AB, Canada,  T6G 2G7 \\
${}^{b}$Department of Physics and Astronomy, \\ University of Waterloo, Waterloo, Ontario, Canada, N2L 3G1}


\maketitle
\begin{abstract}
In the simulations of the multi-black holes and merging black holes a larger primary image and a secondary smaller image which looks like an eyebrow and the deformation of the shadows have been observed. However, this kind of eyebrow-like structure was considered as unique feature of multi black hole systems. In this paper, we illustrate the new result that  in the case of octupole distortions of a Schwarzschild black hole the local observer sees two shadows or two images for this single black hole, i.e., also an eyebrow-like structure. Presence of two images in our case is remarkable, as we have only one black hole, however, the observer sees two dark images of this single black hole. 
\end{abstract} 

\normalsize
\baselineskip 17 pt
\newpage

\section{Introduction}
The apparent shape of a black hole, or its shadow, corresponds to the lensed image of the event horizon and  is a two-dimensional dark zone seen by the observer. For a non-rotating Schwarzschild black hole, the shadow is a perfect circle \cite{Synge}. The investigation of the apparent shape of a Schwarzschild black hole \cite{Synge} has been extended to many other cases, including Kerr black holes \cite{11}, (see also \cite{19,20,20b}), Kerr-Newman black holes \cite{12}, Schwarzschild-de Sitter black holes \cite{12B}, Tomimatsu-Sato space-times \cite{13}, black holes in extended Chern-Simons modified gravity \cite{14}, rotating braneworld black holes \cite{15A,15} \footnote{The Randall-Sandrum branewrold black hole solution was constructed in \cite{Toby1,ACP1,ACP2}; it allows for a more extensive study of shadow of static braneworld black holes within the context of Randall-Sandrum model.}, Kaluza-Klein rotating dilaton black holes \cite{16}, charged rotating black holes arising from  heterotic string theory \cite{Hetro}, Kerr-NUT black holes \cite{17}, Kerr-Newman-NUT black holes with a cosmological constant \cite{21}, and multi-black holes \cite{18}. Shadows  for black holes in modified gravity \cite{Moffat}, and the five-dimensional rotating Myers-Perry black hole \cite{Papnoi} have also been considered.

In a recent paper, we considered the shadow of a black hole that is distorted under the effect of external sources of matter \cite{Shad1}.  Distorted black holes approximate those of dynamical black holes that relax on a time scale much shorter than that of the external matter, and have been a subject of long-standing interest  \cite{Geroch, Israel,Chandrasekhar,Tomimatsu,Breton:1997,Peters,Xanthopoulos,Fairhurst,ASF,ASP}.  Defining the concept of a ``local shadow'', a shadow seen by an observer not located at asymptotic infinity, we have investigated how the shadow of a black hole gets deformed under the effect of external matter. The local observer of the shadow is not located at asymptotic infinity, but rather at a finite distance from the black hole,  in a vacuum region that is interior to the external sources. We considered situations where the distorting potential does not modify the undistorted Schwarzschild black hole metric significantly, restricting ourselves to the simplest distortions, namely the quadrupole and octupole distortions of the black hole \cite{Shad1}. 

In the case of multi-black holes, interesting eyebrow-like structures outside the main shadows of the black holes, as well as the deformation of these shadows, have been observed \cite{18}. Such eyebrow structure have also been observed in the case of shadows of two merging black holes \cite{merge,Bohn}. This kind of structure has been regarded as a unique feature of a multi black hole system. Multiple shadows have recently been shown to be present for asymptotically flat rotating black holes with scalar hair that are solutions to  Einstein gravity minimally coupled to a massive complex scalar field \cite{Carlos2015}.

In this paper, we demonstrate that these multiple shadows also occur for single distorted black holes which was pointed out as eyebrow structure in our earlier paper \cite{Shad1}. Specifically, we  demonstrate that  for octupole distortions an observer will see two shadows or two images for a single black hole: a larger primary image and a secondary smaller image that looks like an eyebrow. In the case of two black holes each eyebrow-like shadow corresponds to the rays getting bent by one black hole that eventually get captured by the other black hole. Geodesics traced from one black hole must be deflected around the other black hole to generate an eyebrow, (see figure 3 of \cite{18}, and figure 4 of \cite{Bohn}). Therefore one naturally expects two primary shadows and two eyebrow-like secondary shadows for two black holes. However, the presence of two images in our case is different, as we have only one black hole. For an observer looking at the two dark images of this single black hole, it would appear that  there are two black holes present.

Our paper is organized as follows, in Section 2, we give a brief review the solution representing a distorted axisymmetric Schwarzschild black hole and the equations of motion in the corresponding spacetime. In Section 3, we review the map which enables us to define the local shadow. In Section 4, we present and discuss our results, and finish with Section 5 where, we sum up.  In the Appendix, we find the Newtonian multiple moments associated with our multiple moments. In what follows, we employ units where $G=c=1$.
 
\section{Spacetime and null geodesics}

The metric for an uncharged, non-rotating black hole in an external gravitational field is
\ba\n{St1}
d\mathcal{S}^{2}&=&-\left(1-\frac{2M}{\rho}\right)e^{2U}d\mathcal{T}^{2}+\left(1-\frac{2M}{\rho}\right)^{-1}e^{-2U+2V} d\rho^{2}\nonumber\\
&+&e^{-2U}\rho^{2}(e^{2V}d\theta^{2}+\sin^{2}\theta d\phi^{2})\,,
\ea
where 
\ba
U&=&\sum_{n\ge 0} c_nR^nP_n, \n{Upot}\\
V&=&\sum_{n\ge 1}c_n\sum_{l\ge 0}^{n-1}\left[\cos{\theta}-(2r-1)-(-1)^{n-l}((2r-1)+\cos{\theta})\right]R^{l}P_l \nonumber\\
&+& \sum_{n,k\ge 1}\frac{nkc_nc_k}{n+k}R^{n+k}(P_nP_k-P_{n-1}P_{k-1})\, ,\n{Vpot}
\ea 
where the $c_{n}$'s are the multipole moments and 
\ba
P_n&=&P_n\left((2r-1)\frac{\cos{\theta}}{R}\right) \\
R&=&\sqrt{(2r-1)^2-\sin^2{\theta}} 
\ea
with the $P_n$'s  denoting Legendre polynomials of the first kind.  While a general analysis of metrics of the form (\ref{St1}) was first discussed by Geroch and Hartle \cite{Geroch}, and a prescription for obtaining ${U}={U}(\rho,\theta)$ and $V=V(\rho,\theta)$ was given by Chandrasekhar \cite{Chandrasekhar}, the explicit form of (\ref{Upot}) and (\ref{Vpot}) was first obtained by Breton et.al. \cite{Breton:1997} (albeit in different coordinates
where $x=r/M-1$, and $y=\cos\theta$).

The metric (\ref{St1}) is static and axisymmetric, and is related to the following metric by a combination of conformal transformations and redefinitions of the time coordinate
\ba
ds^2& =& -(1-\frac{1}{r})e^{2\mathcal{U}}dt^2 +{(1-\frac{1}{r})}^{-1}{e^{-2\mathcal{U}+2V}}dr^2 \nonumber\\
&&+e^{-2\mathcal{U}}r^{2}(e^{2V}d\theta^2+\sin^2{\theta}d\phi^2) \,,\n{ST4}\\
&&d\mathcal{S}^{2}=4M^2 e^{-2u_0} ds^2\,,~~~~\mathcal{T}=e^{-2u_0} 2M\, t \, ,~~~
\rho=2M r\, .
\ea
where $\mathcal{U} = {U}-u_0$ with  $u_0=\sum_{n\ge 0}c_{2n}$.

The null geodesic equations are conformally invariant. In our investigation of this spacetime we will require
$$
\sum_{n\ge0}c_{2n+1}=0\,, 
$$ which is the condition ensuring the absence of conical singularities.  This  is equivalent to requiring that
the  black hole does not experience a net force on the axis when placed in the external gravitational field \cite{Geroch,Chandrasekhar}. For $U=V=0$, this metric represents a Schwarzschild black hole.

We shall study the effect of distortions on the black hole by considering the simplest cases  \cite{Shad1} of octupole ($c_{1}=-c_{3}$) distortions. 
For the octupole distortions we have
\ba
\hspace{-0.4cm}V& =& A\cos{\theta}+A_0\cos^2{\theta}+A_1\cos^4{\theta}+A_2 \, ,
\nonumber\\
\hspace{-0.4cm}\mathcal{U}&=&-c_1\cos{\theta}\left[r \cos^2{\theta}(20r^2-30r+12)-r(12r^2-18r+8) +\sin^2{\theta}\right]  ,  \label{UV}
\ea 
where  the functions  $A_i =A_i(r,\theta)$ are given  by
\ba
A& =& 2c_1\sin^2{\theta}\left[6r(r-1)+1\right], \nonumber\\
A_0&=& -2{c_1}^2r\sin^2{\theta}\left[-168r^5+504r^4-570r^3+ 300r^2-72r+6\right], \nonumber\\
A_1&=& -2{c_1}^2r\sin^2{\theta}\left[300r^5-900r^4+1008r^3- 516r^2+117r-9\right], \nonumber\\
A_2&=& -2{c_1}^2r\sin^2{\theta}\left[12r^5-36r^4+42r^3-24r^2+7r-1\right] . \label{Afunctions}
\ea
In the appendix, we show that the parameters $c_n$ are related to the Newtonian multiple moments $\tilde{c}_n$ of a static axisymmetric mass distributions whose interior gravitational field is described by a vacuum solution. This relationship to  the Newtonian potential is useful in  understanding the effects of the multipole moments $c_n$.  As an example, consider two masses $m_1$ and $m_2$, located at distances $\rho_1$ on the upper plane and $\rho_2$ on the lower one, from the centre, on the axis, respectively, and a thin ring of mass $m$ and radius $a$ on the equator. Comparing to this Newtonian case, we can interpret the different multiple moments $c_n$ in terms of the $\tilde{c}_n$.

 The case of only even multiple moments, which in the light of the fact that ${\tilde c}_2>{\tilde c}_4>... $, can be considered as approximately ${\tilde c}_2\neq0$ and all other even multiple moments ${\tilde c}_{2n}=0$ (for $n>1$), can be achieved by a ring of mass $m$ around the black hole on the equatorial plane. In the Newtonian potential, the case of ${\tilde c}_3=-{\tilde c}_1\neq 0$ and ${\tilde c}_2=0$, and $c_{n}\sim0$, ($\forall~ n>3)$ can be achieved with two mass sources $m_1>0$ and $m_2>0$, located on the upper and on the lower axis and a thin ring of mass $m>0$ around on the equatorial plane. Note that we are taking into account the fact that higher order multiple moments are much smaller than the lower order multiple moments. Therefore we conclude that  it is also possible to have ${ c}_3=-{ c}_1\neq 0$ and ${ c}_2=0$, and $c_{n}\sim0$, ($\forall~ n>3)$ with a similar configuration of mass sources, as an example, with only positive masses.

The source of the multipole moments need to correspond to reasonable distributions of matter. Here, we consider the distorted black hole as a local solution, valid only in a certain neighbourhood of the horizon. 
The external sources distorting the black hole are located beyond this neighbourhood where the spacetime is not vacuum and the solution (\ref{ST4}) is not valid. In other words, even though our metric represents a vacuum space-time, some matter sources exist exterior to the region of validity of the solution and cause distortion of the black hole. A global solution can be constructed extending the metric to an asymptotically flat solution by a sewing technique. This can be realized by cutting the spacetime manifold in the region where the metric is valid and attaching to it another spacetime manifold where the solution is not vacuum anymore, but the sources of the distorting matter are also included. Beyond this non-vacuum region, we  assume to have an asymptotically flat vacuum solution.  In the presence of sources, the $\{tt\}$ component of the Einstein equations reads 
\ba
R_{\alpha\beta} \delta^\alpha_t \delta^\alpha_t&=& 8\pi(T_{\alpha\beta}- \frac{T^\gamma_{~\gamma}}{3}g_{\alpha\beta}) \delta^\alpha_t \delta^\alpha_t\nonumber\\
&=&\frac{(r-2M)}{r^3} e^{(4U-2V)} \triangle U \,, \n{Tenergy}
\ea where $R_{\alpha\beta}$ is the Ricci tensor, $T_{\alpha\beta}$ is the energy-momentum tensor representing the sources, $\triangle$ is the Laplace operator. If the sources satisfy the strong energy condition, the right hand side of (\ref{Tenergy}) must be non-negative. Since the Laplace operator $\triangle$ is a negative operator, the strong energy condition implies that $U<0$, assuming $U=0$ at asymptotic infinity. Considering the explicit form of $U$ for octupole distortions, we can see that with an appropriate value of $c_0$, the function $U$ can be negative in the region of interest, which is the sewing radius. Although we are not considering $c_0$ in our calculations,  we assume that it has an appropriate value.

The null geodesic equations  
\be
\frac{d^2 x^\mu}{d\tau^2}+\Gamma_{\alpha\nu}^\mu \frac{d x^\alpha}{d\tau}
\frac{d x^\nu}{d\tau}=0,
\ee 
with  $\Gamma_{\alpha\nu}^\mu$  the Christoffel symbols,  become
\ba
r''&=&-\frac{r'^{2}}{2}\left[\frac{h_{,r}}{h}+2\frac{f_{,r}}{f}+\frac{3-4r}{r(r-1)}\right]+\frac{\theta'^{2}}{2}(r-1)\left[r \frac{h_{,r}}{h}+2\right]\nonumber\\
&+& \theta' r' \left[-\frac{f_{,\theta}}{f}-\frac{h_{,\theta}}{h}+2\cot\theta\right]-\frac{r^{3}\sin^{4}\theta}{2l_{z}^{2}hf^{3}(r-1)}\left[1+r(r-1)\frac{f_{,r}}{f}\right]\nonumber\\
&-&\frac{1}{hf}\sin^{2}\theta (r-1)\left[\frac{rf_{,r}}{2f}-1\right],\\
\theta''&=&\frac{r'^{2}h_{,\theta}}{2r(r-1)h}
+\frac{\theta'^{2}}{2}\left[- \frac{h_{,\theta}}{h}-2\frac{f_{,\theta}}{f}+4\cot\theta\right]-\theta' r' \left[\frac{h_{,r}}{h}+\frac{f_{,r}}{f}\right]\nonumber\\
&-&\frac{r^{3}\sin^{4}\theta f_{,\theta}}{2l_{z}^{2}hf^{4}(r-1)} - \frac{\sin^{2}\theta}{2hf}\left[\frac{f_{,\theta}}{2f}-2\cot\theta\right]\,, 
\ea 
for the spacetime (\ref{ST4}), 
where $f=e^{2\mathcal{U}}$, $h=e^{-2\mathcal{U}+2V}$, $r''=d^{2}r/d\phi^{2}$, 
$\theta''=d^{2}\theta/d\phi^{2}$. The prime denotes a derivative with respect to $\phi$ and an overdot denotes a derivative with respect to the proper time $\tau$: 
\ba
\dot{r}&=&\frac{dr}{d\tau}=\frac{dr}{d\phi}\frac{d\phi}{d\tau}=r'\dot{\phi}\, , \nonumber\\
\dot{\theta}&=&\frac{d\theta}{d\tau}=\frac{d\theta}{d\phi}\frac{d\phi}{d\tau}=\theta'\dot{\phi} \, . \n{devchange}
\ea 
Moreover, since $\partial/\partial t$ and  $\partial/\partial \phi$ are both Killing vectors, the quantities $E$ and $L_z$ are conserved quantities along geodesics of the space-time 
\ba
E &\equiv&-u_{\mu}\xi^{\mu}_{(t)}=\left(1-\frac{1}{r}\right)e^{2\mathcal{U}}\,\dot{t}\, ,\nonumber\\
L_{z} &\equiv&u_{\mu}\xi^{\mu}_{(\phi)}=e^{-2\mathcal{U}}r^{2}\sin^{2}\theta\ \dot{\phi} \, ,\n{Con}
\ea 
where $l_{z}={L_{z}/E}$.  Finally, we also have the following constraint
\be\n{constraint}
u_{\mu}u^{\mu}=0\, \Rightarrow
 -\frac{r^4\sin^4\theta}{f^2l_z^2}+r^2\sin^2\theta \, F+fhr'^2+r^2Ffh\theta'^2=0 \,, 
\ee
where  $F=1-1/r$. 

\section{Local Shadow}

Consider a photon that gets emitted from a point located on a sphere of radius ${r}_{e}$. Following the motion of this null ray forward, it will either get captured by the black hole or reach the eye of an observer located at $(r_{o},\theta_{o})$, with $\theta_{o}$ the polar angle from the axis of symmetry.  This observer is assumed to be located in the interior region, which means that the sources responsible for the distortion of the black hole are all located at $r > r_{o} \geq r_{e}$. To determine the shadow of the black hole, we instead trace the trajectory of these photons backward i.e., the initial point of motion of the photons is at $(r_{o},\theta_{o})$.

Suppose  the observer is looking in the direction of the centre of the black hole. The tangent vector to a null ray at the observation point $(r_{o},\theta_{o})$ can be written as
\be
u^{\mu}=\alpha(-e^{\mu}_{(t)}+\bar{w} \, e^{\mu}_{(r)}+\bar{p} \, e^{\mu}_{(\theta)}+\bar{q} \,  e^{\mu}_{(\phi)}),\n{tangent2}
\ee
where $\bar{w}$, $\bar{p}$, and $\bar{q}$ are displacement angles, $e^{\mu}_{(m)}$'s are the orthonormal vectors in the observation point and $\alpha$ is a scalar coefficient. Tracing straight backward null rays leaving the observer at  angles $(\bar{p}$, $\bar{q})$, as if the space were flat, it appears to him/her that the photon has reached his/her eye from the point ${p}=\bar{p} r_o$ and $q=\bar{q} r_o$ on the plane of the black hole. The initial condition of such a ray, i.e., its $l_z$ and ${\theta'}_{o}$ are given in terms of the two angles $(\bar{p}, \bar{q})$ or $(p=r_o \, \bar{p}, \,q=r_o \, \bar{q})$,
\be
p=r_o \,  \bar{p}=\pm\frac{l_{z}}{\sin^{2}\theta}f^{\frac{3}{2}}\sqrt{h(1-\frac{1}{r})}\theta'\Bigg|_{o} \qquad 
q=r_o \, \bar{q}=-\frac{l_{z}}{\sin\theta}f\sqrt{1-\frac{1}{r}}\Bigg|_{o} .  \n{t2a}
\ee The ``local shadow'' is defined in terms of two angles $(\bar{p}, \bar{q})$ or $(p=r_o \, \bar{p}, \,q=r_o \, \bar{q})$ for an observer located at $(r_{o},\theta_{o})$, who is tracing back the rays. If a null ray with initial $(p, q)$ is absorbed by the black hole, we consider the corresponding value of $(p, q)$ to be a black point on the plane of the black hole -- by definition it is  a member of the ``local shadow'' of the black hole. The map  (\ref{t2a}) for an observer at infinity (with $f=h=1$ ) matches with the equations for the map of undistorted black holes \cite{11}.

For our numerical calculation, we choose the values of $p$, $q$ and $(r_o,\theta_o)$ as  initial conditions. For every photon trajectory with initial values $p$ and $q$, we compute $l_z$, ${\theta'}_{o}$ at $(r_o,\theta_o)$. Furthermore, $r'$ at $(r_o,\theta_o)$ is computed from \eq{constraint} and we derive
\be
 r'(r_{o},\theta_{o}) =   \left(\frac{r^2 \sin^4\theta}{f^3 l_z^2 h}\right)^{\frac{1}{2}}\Big|_{(r_{o},\theta_{o})}\sqrt{r^2_{o} -p^2-q^2} 
\n{xy}
\ee 
using \eq{t2a}.   We see from \eq{xy} that only those satisfying the condition $p^2+q^2\le {r_{o}}^{2}$ are acceptable. 

Numerically, we divide the available range of $(p,q)$, (a circle of radius $r_o$), to small pixels of size 0.01 and in some cases 0.001. We compute the corresponding photon trajectory for each of these values and determine whether or not it escapes the black hole, namely reaches the radius $r_{e}$. 
For all multipole moments we require for each trajectory that the maximum deviation
from the constraint (\ref{constraint})  is always less than $10^{-3}$. The smallest value of $q$ is $0.01$.
 
For the particular case of octupole distortion  we observe the following symmetry 
\be
V(r,\theta,c_1)=V(r,\pi-\theta,-c_1), ~~~\mathcal{U}(r,\theta,c_1)=\mathcal{U}(r,\pi-\theta,-c_1)
\ee 
 from the  \eq{UV} and \eq{Afunctions}. The spacetime metric is symmetric with respect to the transformation $(\theta,c_1)\rightarrow (\pi-\theta,-c_1)$ and so for an observer located on the equatorial plane (as we can see from \eq{t2a})  the shape of the shadow for  positive and negative values of the multiple moment $c_1$ are related to each other by the transformation 
\be
c_1\rightarrow -c_1,~~~ p\rightarrow -p. 
\ee

\section{Results and Discussion}

We study the eyebrow structure in the case of octupole distortion, for which $c_3=-c_1$. Our previous work \cite{Shad1} indicated that one may observe an eyebrow-like structure (or secondary shadow) in the case of a single distorted black hole. In what follows, we call the larger shadow the primary image and we refer to the smaller one, which is present only for $|c_{3}|\ge|c_{crit}|=1/998$, as the secondary image or eyebrow. 

In this section, we compare the distorted spacetime to undistorted one, presenting  the shapes of the primary and the secondary images of the shadow of the black hole distorted by an octupole parameter. We present the initial conditions, (${\theta'}_{o}  ,  l_{z} $) and graphs for the photon trajectories that correspond to the primary and secondary images of the shadow.
 \begin{figure}[htp]
\setlength{\tabcolsep}{ 0 pt }{\scriptsize\tt
		\begin{tabular}{ cc }
	\includegraphics[width=7 cm]{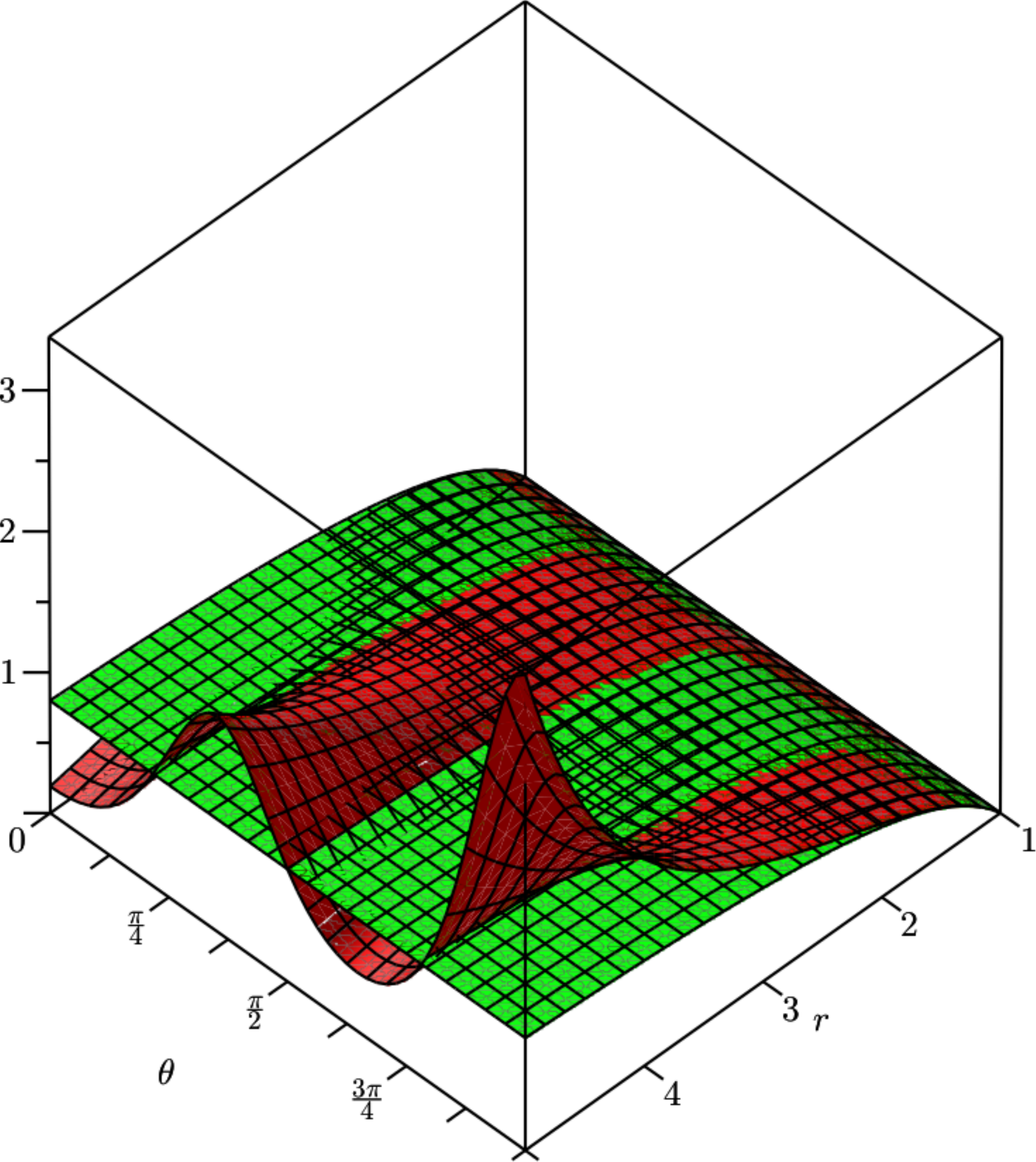}\hspace{2cm} &
            \includegraphics[width=7 cm]{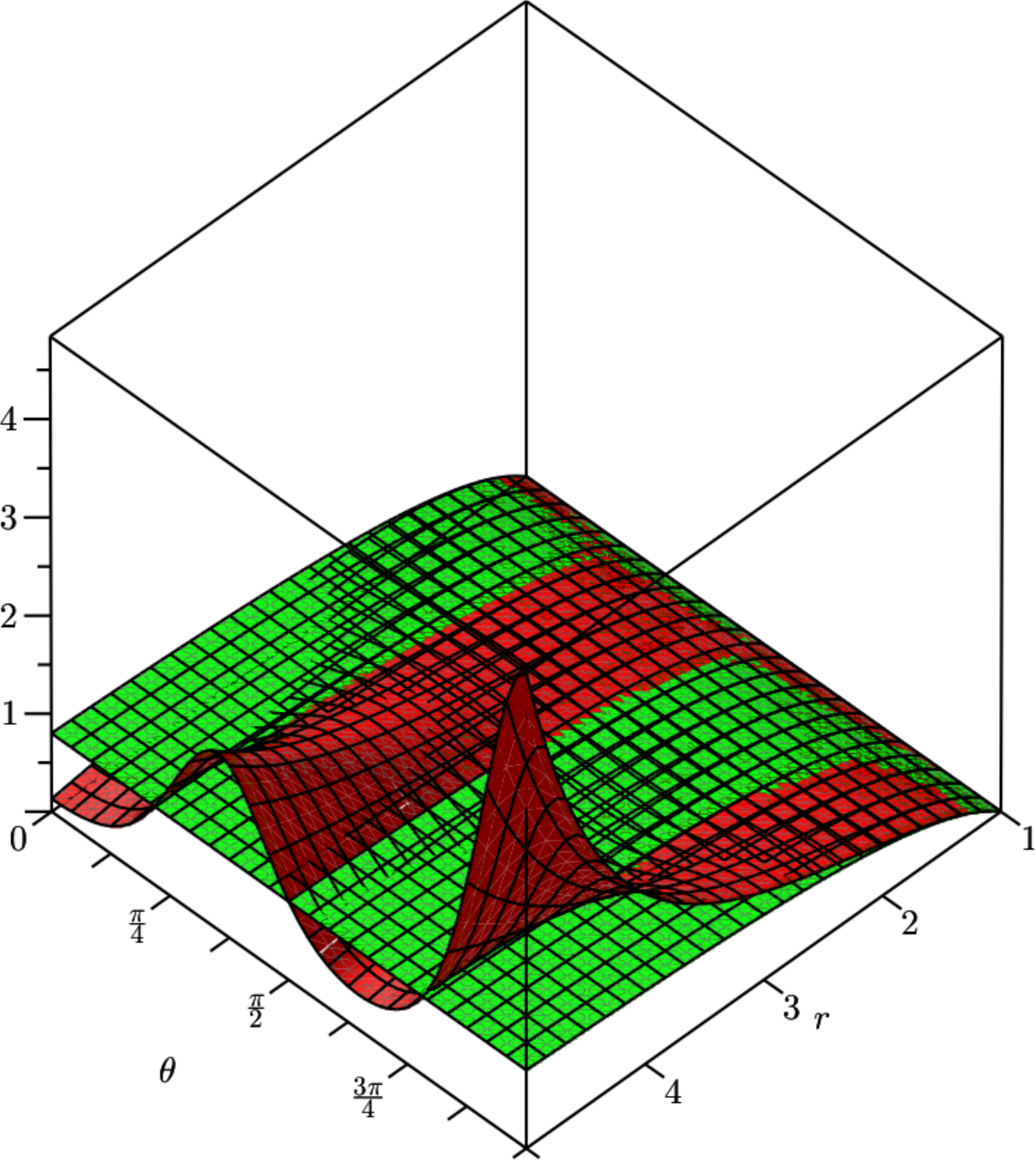} 
  \end{tabular}}
         \caption{Comparison of $g_{ttd}$ of the distorted black hole (red), to  $g_{tt}$ of the Schwarzschild black hole (green). 
  Left: $c_3=-1/1000$; 
  right for $c_3=-1/800$. }\label{f1}
\end{figure}
\begin{figure}[htp]
\setlength{\tabcolsep}{ 0 pt }{\scriptsize\tt
		\begin{tabular}{ cc }
	\includegraphics[width=7 cm]{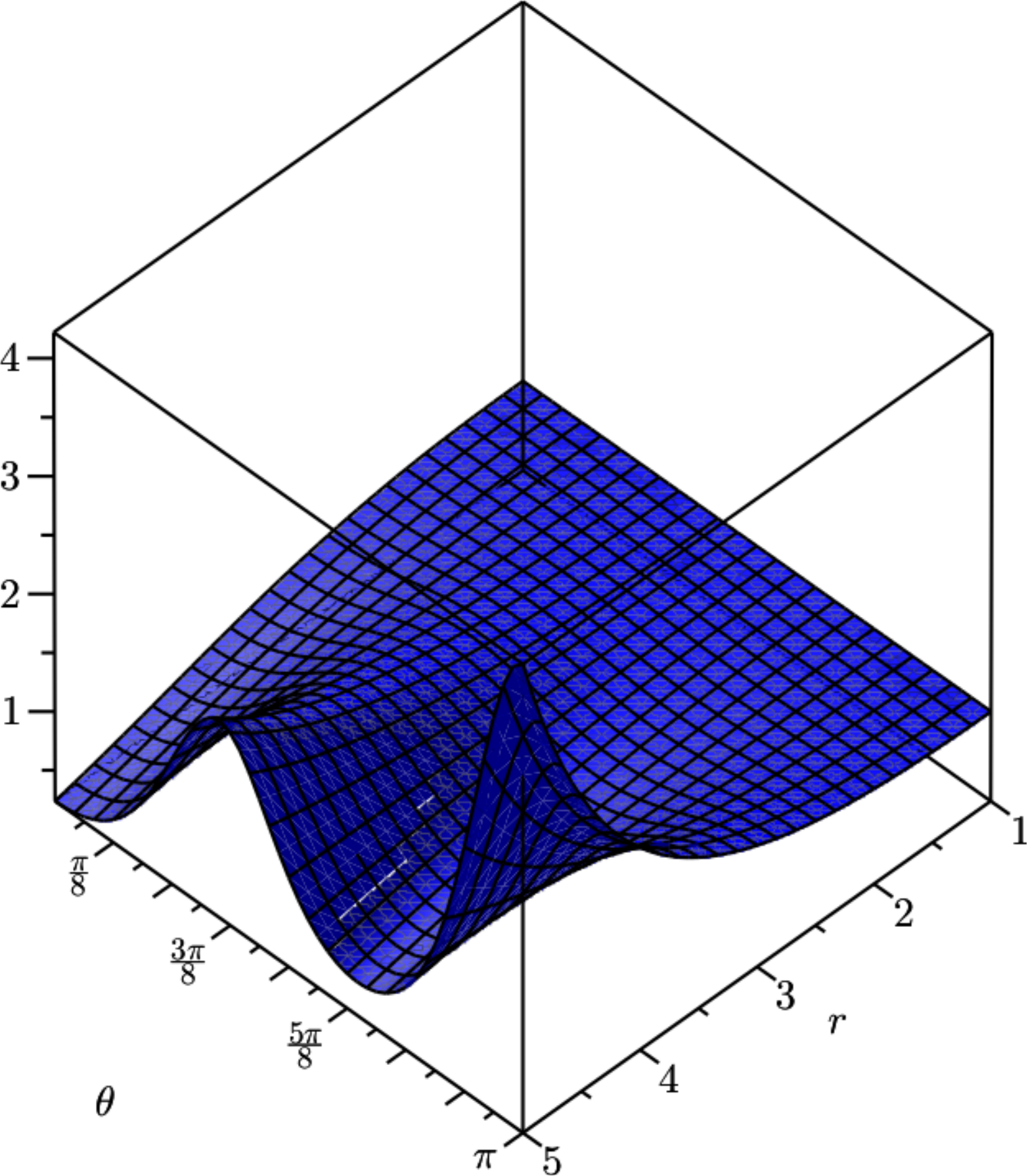}\hspace{2cm} &
            \includegraphics[width=7 cm]{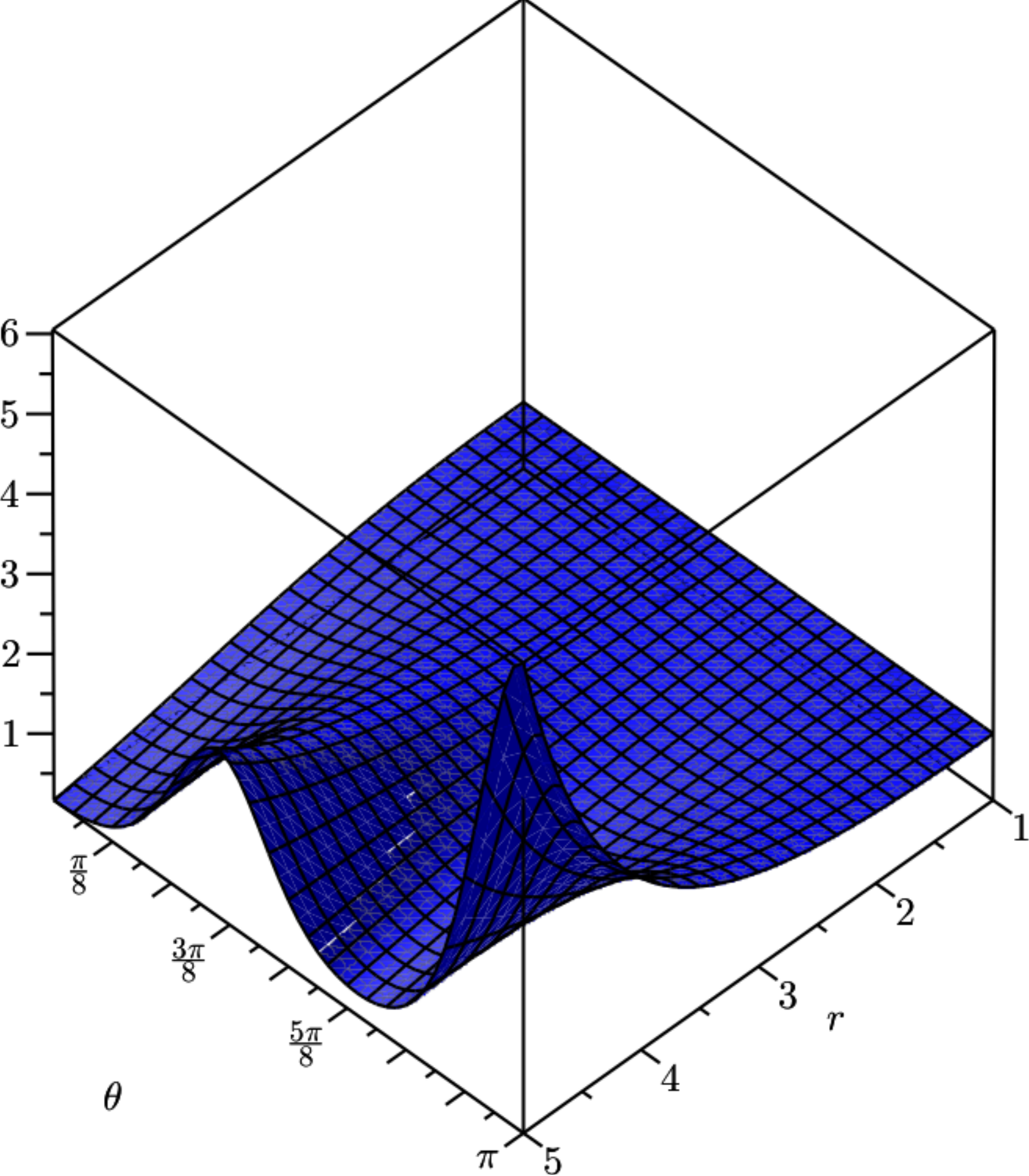} 
  \end{tabular}}
         \caption{Ratio of $g_{ttd}/g_{tt}$.
  Left: $c_3=-1/1000$; 
  right for $c_3=-1/800$. }\label{1}
\end{figure} 

Let us first compare the distorted spacetime to undistorted one. For the octupole distortion, we choose  values of the multiple moment $c_3=-c_1$ such that the $g_{tt(d)}$ (the $tt$-component of the distorted black hole metric) is not very different from the $tt$-component of the undistorted Schwarzschild black hole metric,  namely $f=g_{tt(d)}/g_{tt}<10$. Figure \ref{f1}, depicts the $tt$-components of the distorted (red) and undistorted (green) metrics, and figure \ref{1} illustrates their ratio $g_{ttd}/g_{tt}$ for $c_3=-1/1000$ and $c_3=-1/800$.

\begin{figure}[htp]
\setlength{\tabcolsep}{ 0 pt }{\scriptsize\tt
		\begin{tabular}{ cc }
	\includegraphics[width=7 cm]{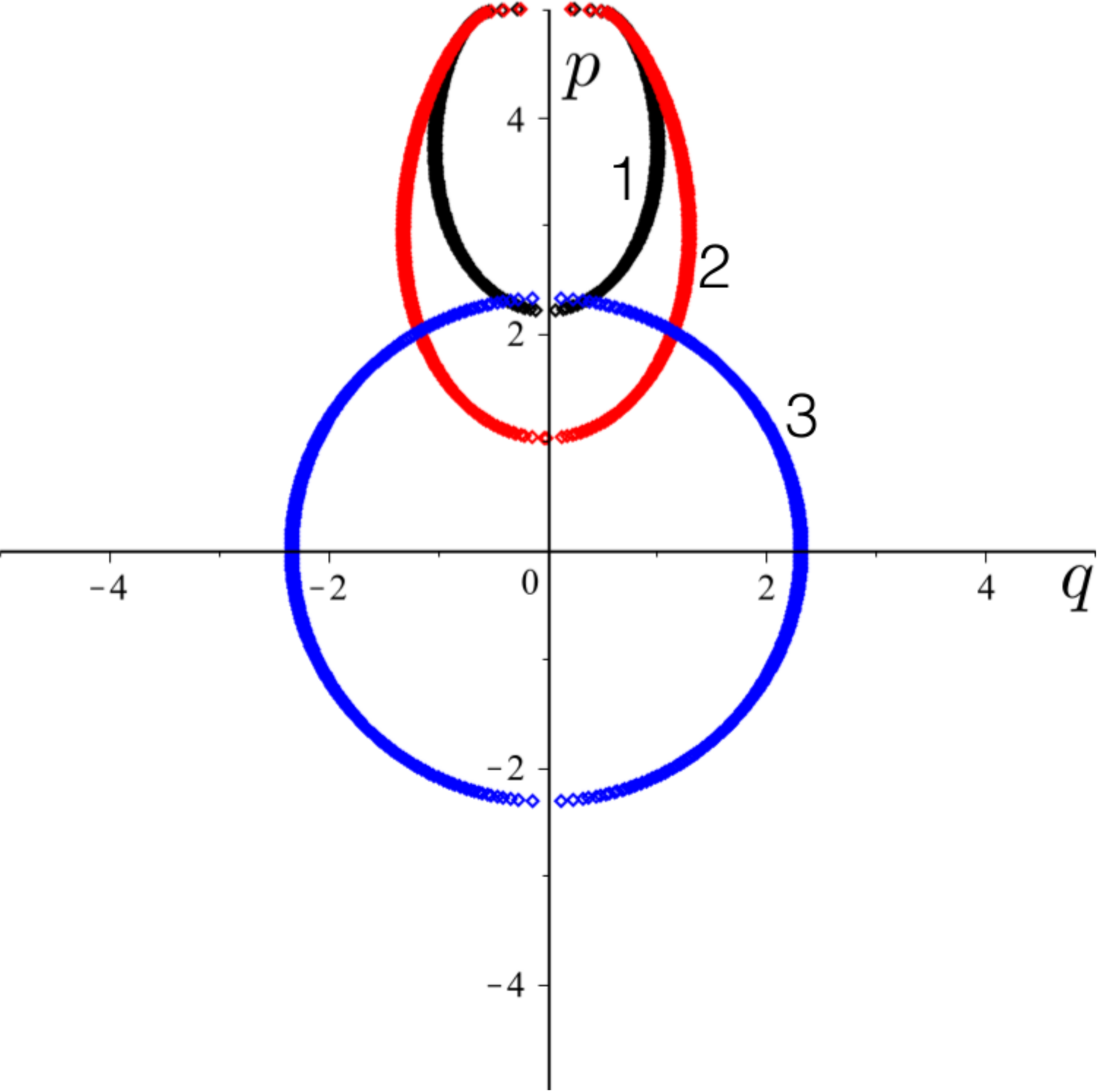}\hspace{2cm} &
            \includegraphics[width=7 cm]{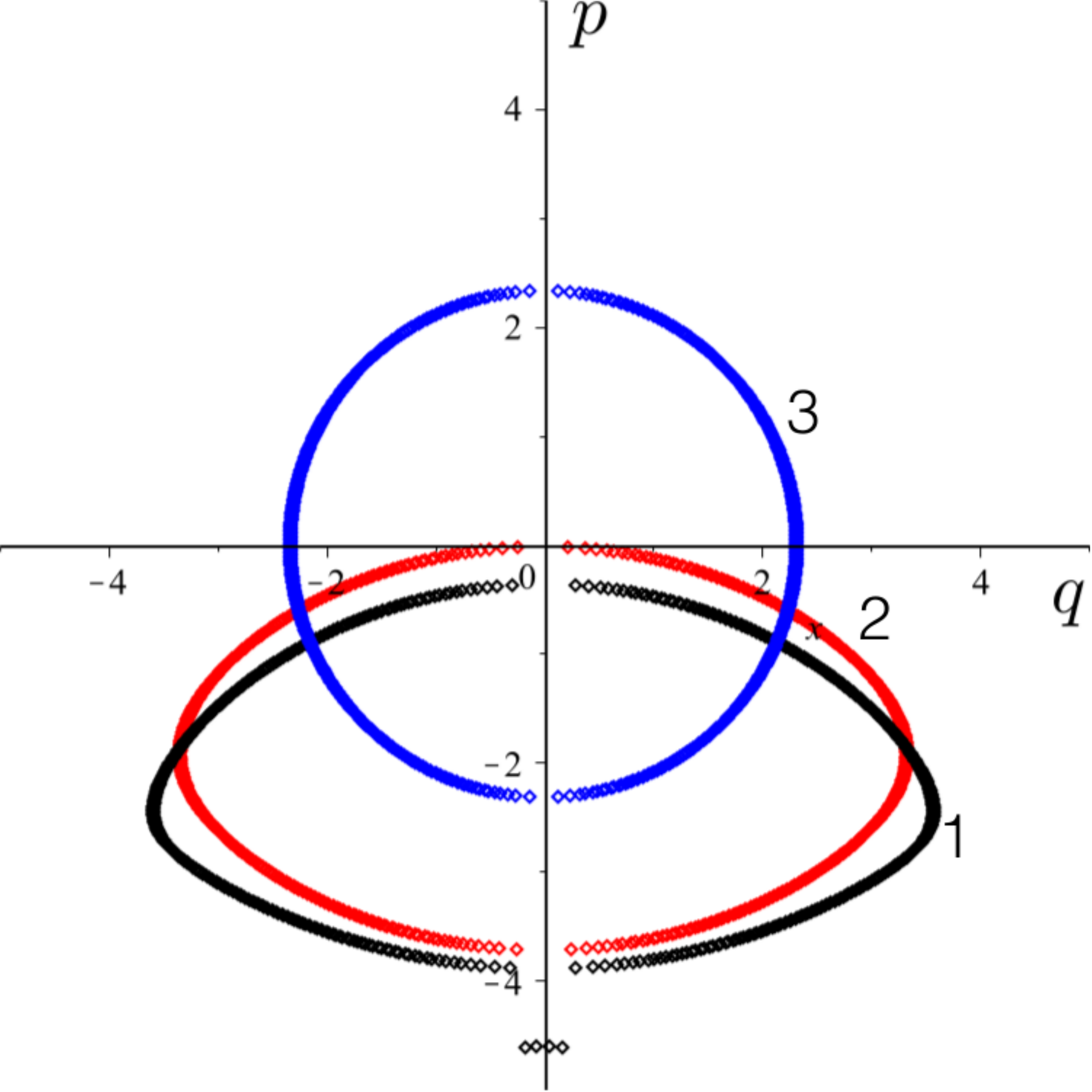} 
  \end{tabular}}
         \caption{Two depictions of the shadow of a black hole for an observer at $\theta_o=\pi/4$ and radius $r_o=5$. Left: the values of the multiple moments are $c_3=\frac{1}{800}$, (with black/line 1), $c_3=\frac{1}{1000}$, (with red/line 2) and $c_3=0$ (which is the undistorted case with blue/line 3). Right:   the values of the multiple moments are $c_3=-\frac{1}{800}$, (with black/line 1), $c_3=-\frac{1}{1000}$, (with red/line 2) and $c_3=0$ (which is the undistorted case with blue/line 3). Note the shadow near $p=-4.5$ for $c_3=-\frac{1}{800}$. }\label{f2}
\end{figure}

\begin{figure}[htp]
\setlength{\tabcolsep}{ 0 pt }{\scriptsize\tt
		\begin{tabular}{ cc }
	\includegraphics[width=7 cm]{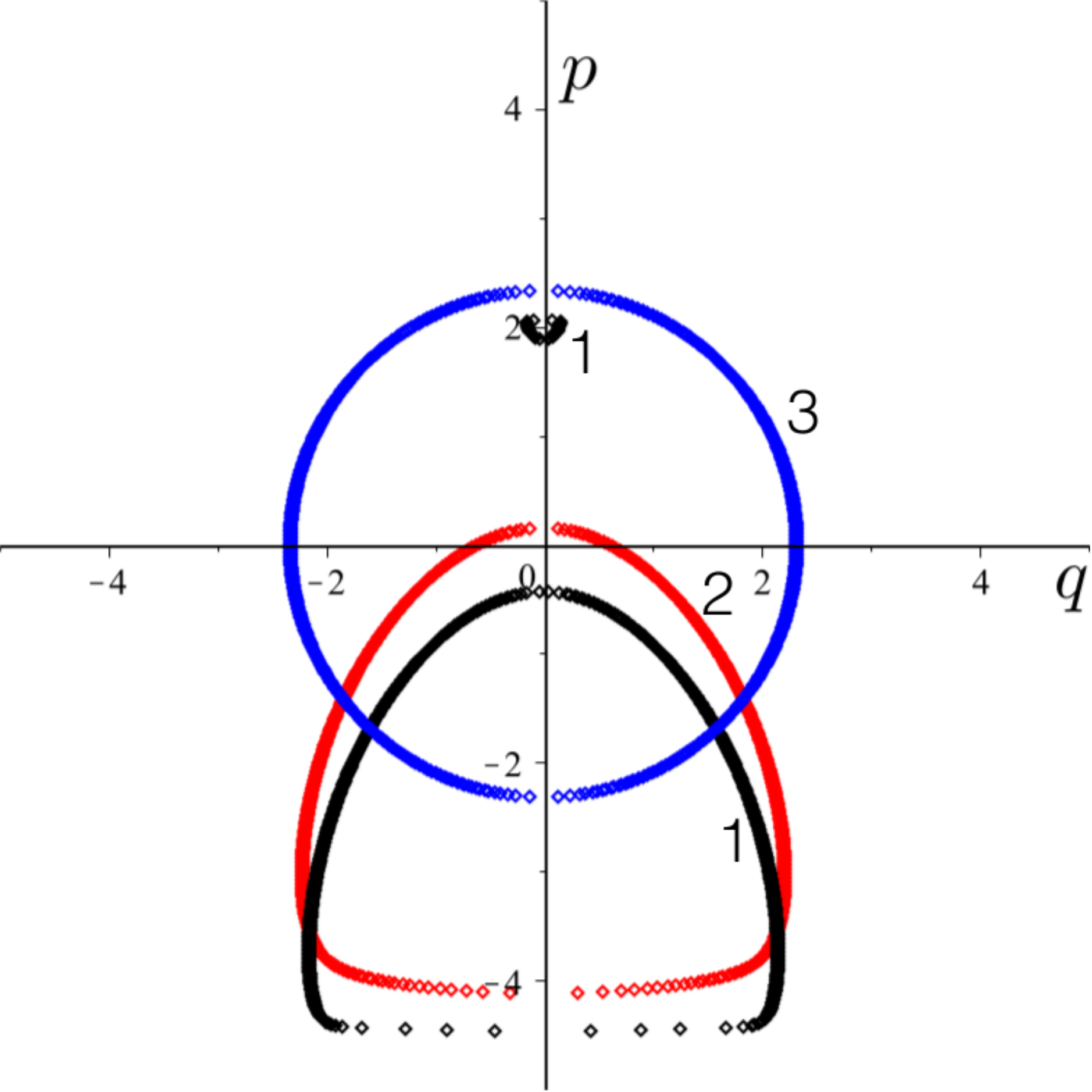}\hspace{2cm} &
            \includegraphics[width=7 cm]{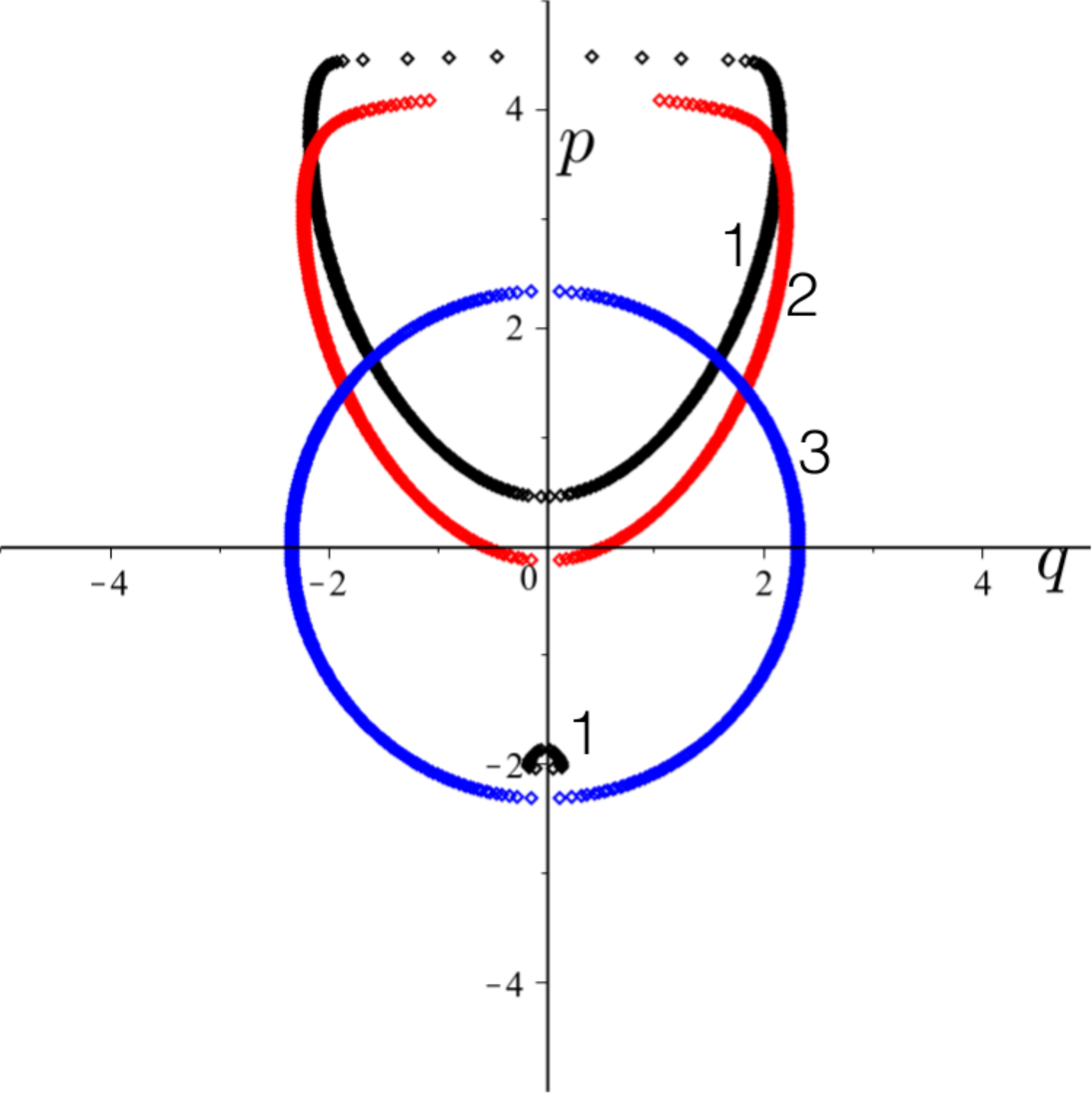} 
  \end{tabular}}
         \caption{Depiction of the shadow of a black hole for an observer at $\theta_o=\pi/2$ and radius $r_o=5$. The values of the multiple moments are (left) $c_3=\frac{1}{800}$ (with black/line 1), $c_3=\frac{1}{1000}$ (with red/line 2), and $c_3=0$, (which is the undistorted case with blue/line 3), and (right)  $c_3=-\frac{1}{800}$, (with black/line 1) $c_3=-\frac{1}{1000}$, (with red/line 2) and $c_3=0$, (with blue/line 3). Note the shadows near $p=\pm 2$ for $|c_3|=\frac{1}{800}$.} \label{f3}
\end{figure}

\begin{figure}[htp]
\setlength{\tabcolsep}{ 0 pt }{\scriptsize\tt
		\begin{tabular}{ cc }
	\includegraphics[width=6 cm]{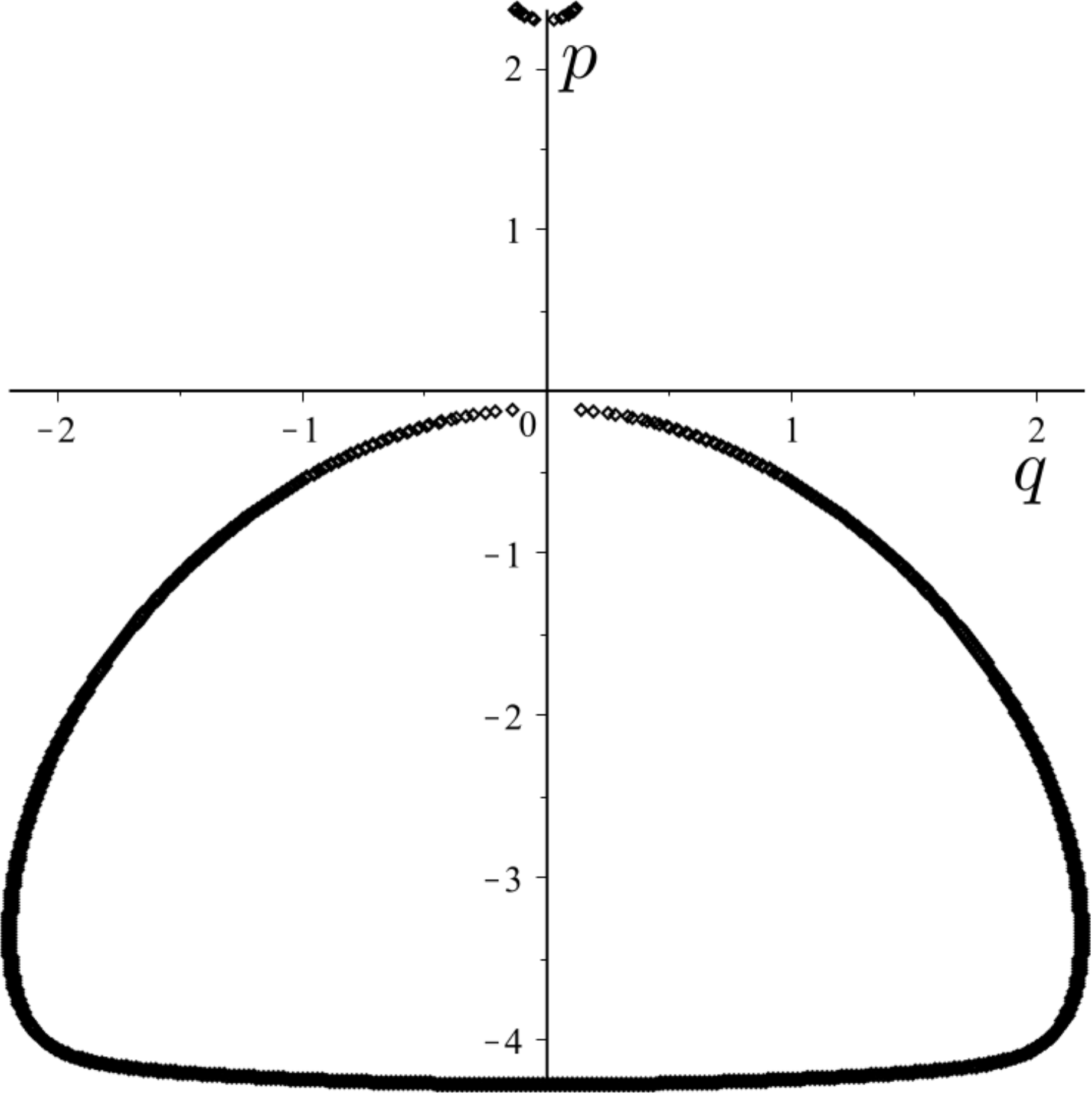}\hspace{3cm} &
            \includegraphics[width=6 cm]{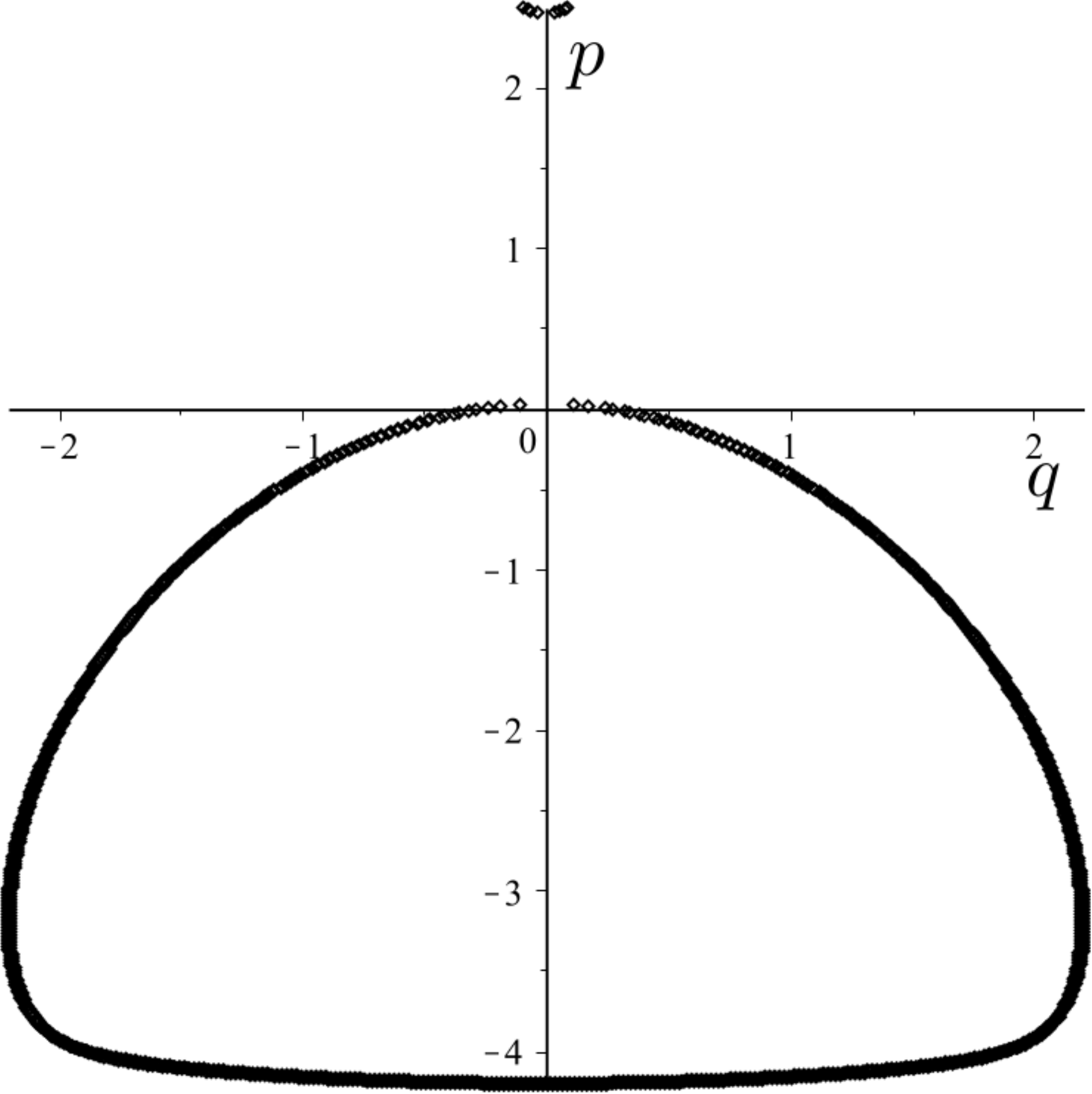} 
  \end{tabular}}
         \caption{Illustration of the shadow of a black hole for an observer at $\theta_o=\pi/2$ and radius $r_o=5$. The values of the multiple moments are (left) $c_3=\frac{1}{900}$ and (right) $c_3=\frac{1}{950}$.}\label{f4}
\end{figure}

In Figure \ref{f2}, we delineate the shadow for a distorted black hole, for positive and negative values of $c_3$, as seen by an observer located at $(r_o=5,\theta=\pi/4)$, providing comparison with the undistorted case. For $c_3=-1/800$, a very small eyebrow structure (or second shadow ) is observed.  Analogous results for  $(r_o=5,\theta=\pi/2)$ are illustrated in figure \ref{f3}. Eyebrow structures are observed for both positive and negative values of $c_3$, ($c_3=-1/800$ and $c_3=1/800$). 
In figure \ref{f4}, we can see the shadow of the distorted black hole for an observer in the equatorial plane, for $c_3=1/900$ and $c_3=1/950$. From figures \ref{f3} and \ref{f4}, we see that there is a secondary image of the same black hole. 

\begin{figure}[htp]
\setlength{\tabcolsep}{ 0 pt }{\scriptsize\tt
		\begin{tabular}{ cc }
	\includegraphics[width=6 cm]{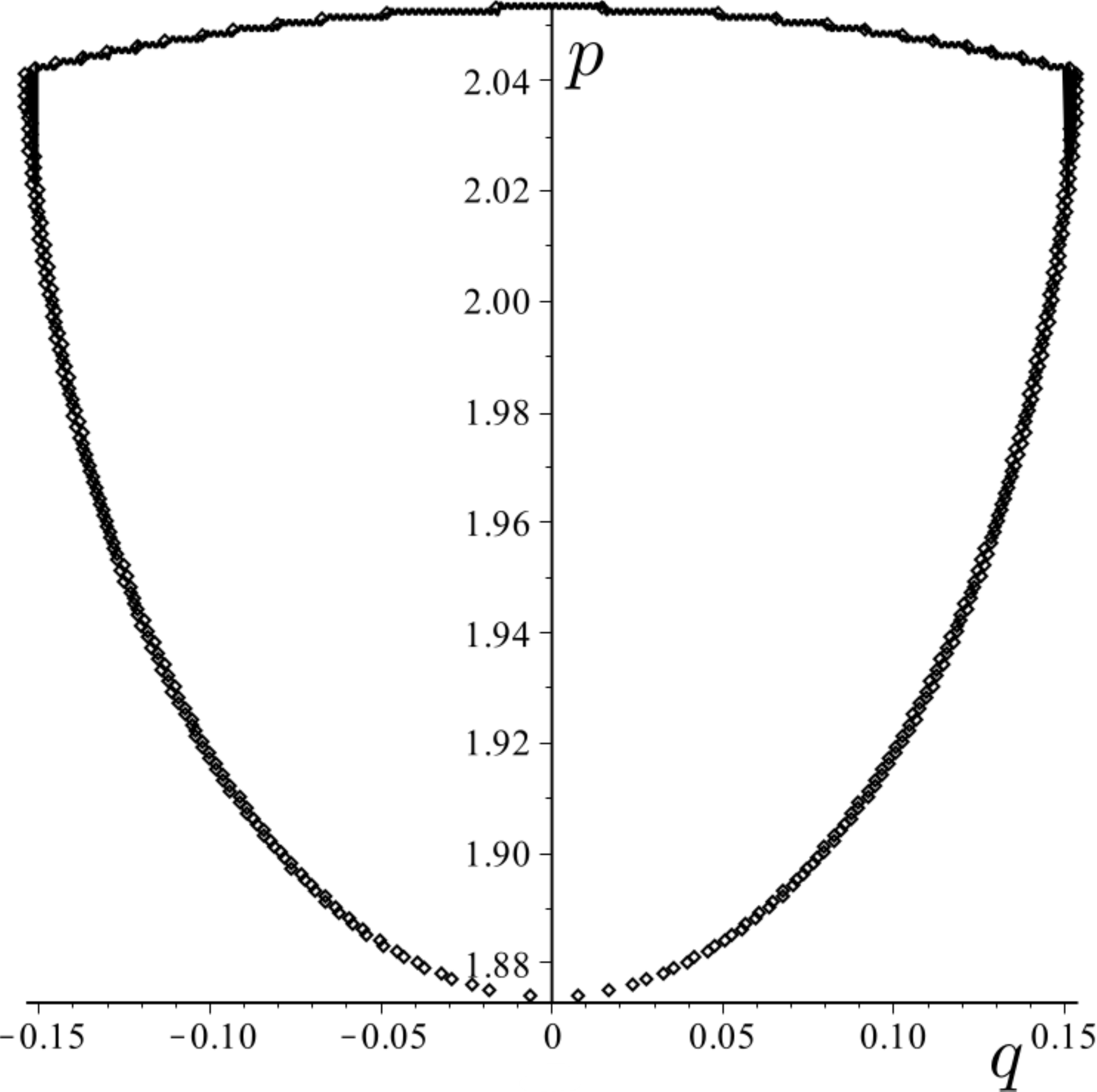} \hspace{3cm} &
              \includegraphics[width=6 cm]{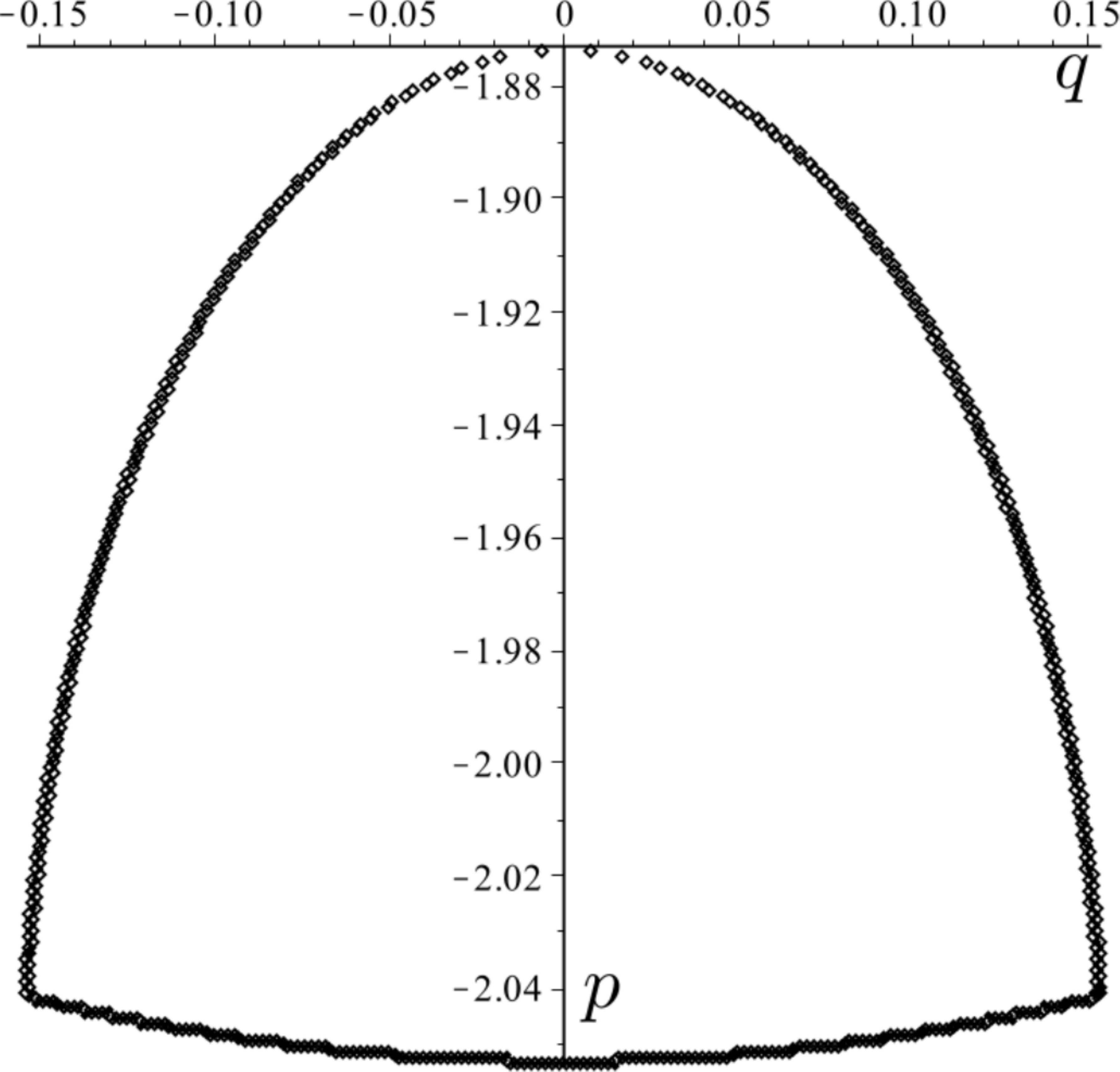} \\ 
  \end{tabular}}
         \caption{Zoom to the eyebrow structure for case of $c_3=1/800$ on the left and on the right $c_3=-1/800$. The observer is located on the equator and $r_o=5$.}\label{f5}
\end{figure}

\begin{figure}[htp]
\setlength{\tabcolsep}{ 0 pt }{\scriptsize\tt
		\begin{tabular}{ cc }
		\includegraphics[width=6 cm]{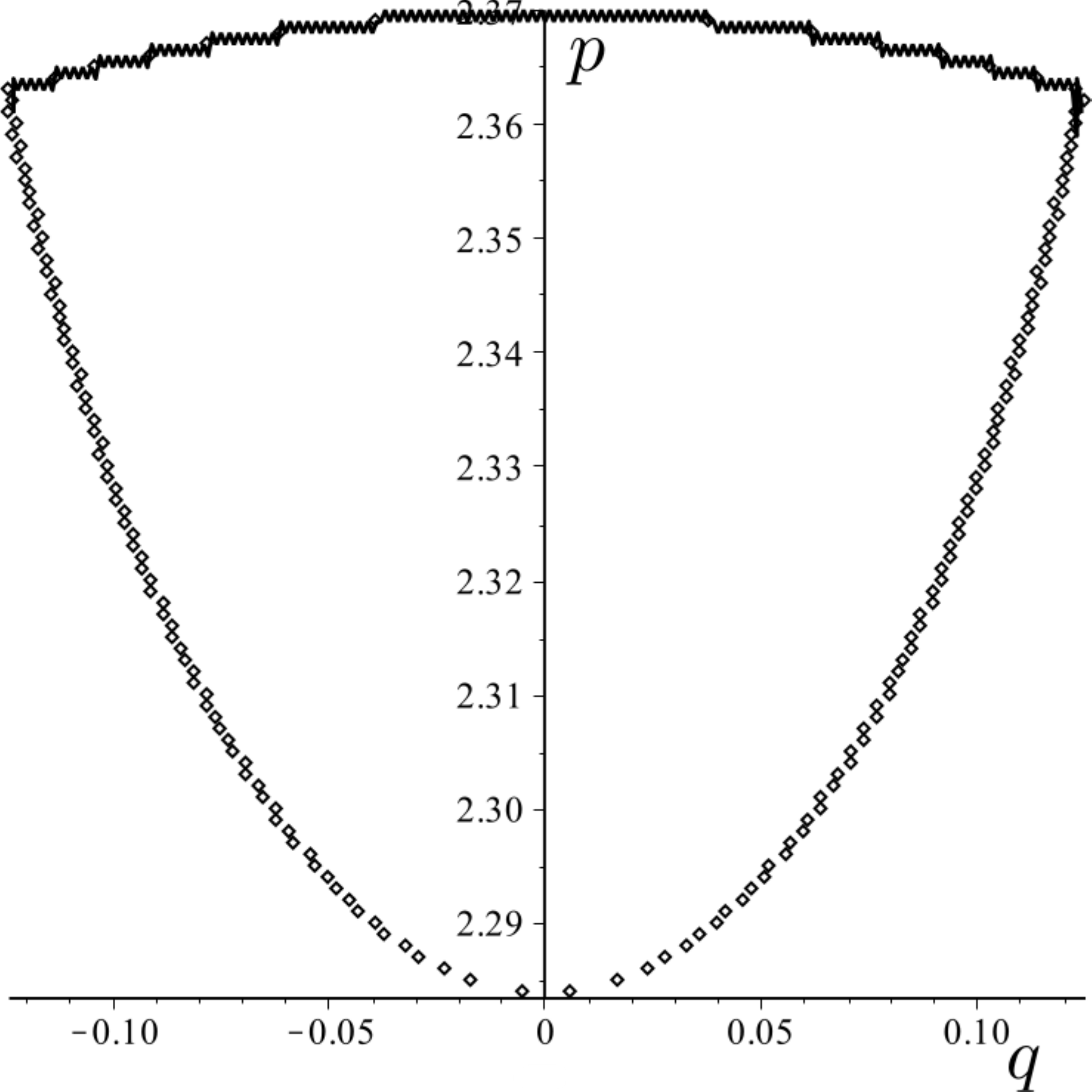} \hspace{3cm} &
	 \includegraphics[width=6 cm]{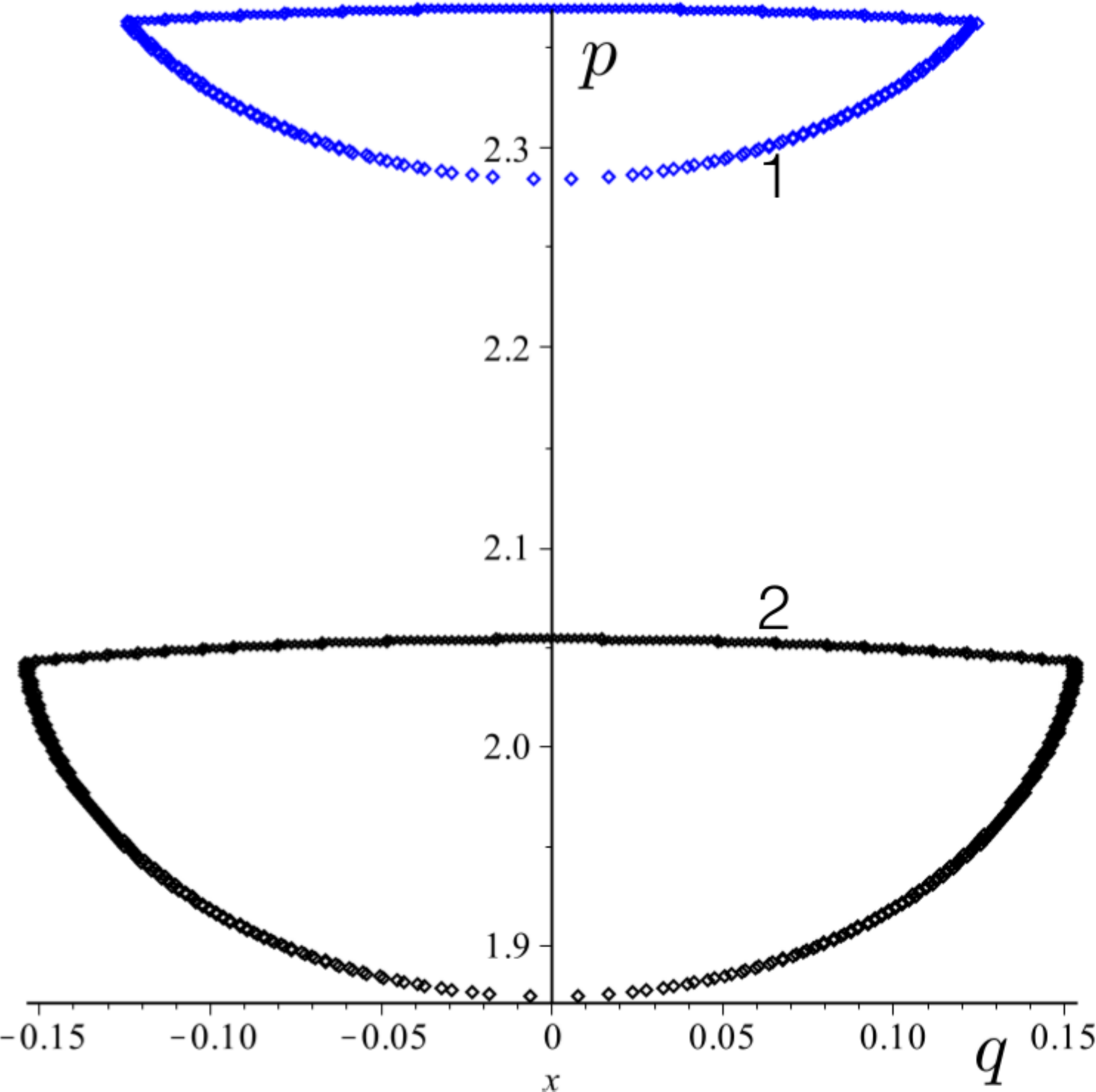} \\
  \end{tabular}}
         \caption{Zoom to the eyebrow structure for case of $c_3=1/900$ on the left and on the right we see the comparison between the cases of $c_3=1/900$ (with blue colour/line 1) and $c_3=1/800$ (with black colour/line 2). The observer is located on the equator and $r_o=5$.}\label{f6}
\end{figure}

To further investigate these secondary smaller shadows we zoom in closer to the image, dividing the appropriate range of $(p,q)$ into smaller pixels of size $0.001$ rather than $0.01$. We illustrate these `close-up' results in figure \ref{f5}, for which $c_3=1/800$ and $c_3=-1/800$ and figure \ref{f6}, for which   $c_3=1/800$ and $c_3=1/900$.  Figures \ref{f3} and \ref{f5}, show the expected symmetry $c_3\rightarrow -c_3$, and $p\rightarrow -p$ for an observer located at the equatorial plane.

We see from figure \ref{f6}, that as the value of the distortion parameter  gets smaller the secondary image of the same black hole  gets smaller as well. We can see from figure \ref{f2}, for $c_3=-1/1000$ this second shadow does not exist. 

\begin{figure}[htp]
\setlength{\tabcolsep}{ 0 pt }{\scriptsize\tt
		\begin{tabular}{ cc c }
		\includegraphics[width=5 cm]{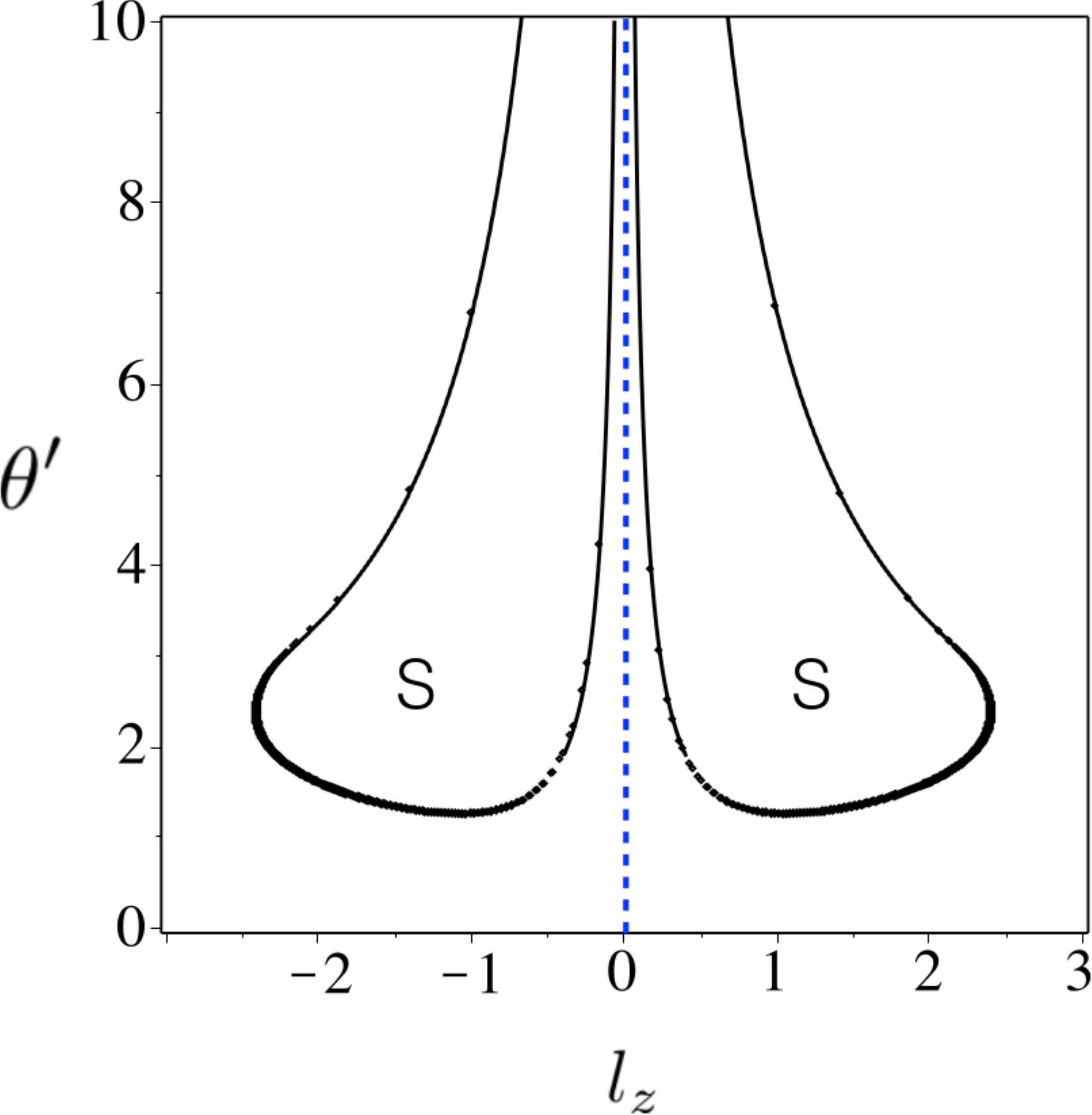} \hspace{0.5cm} &
	 \includegraphics[width=5 cm]{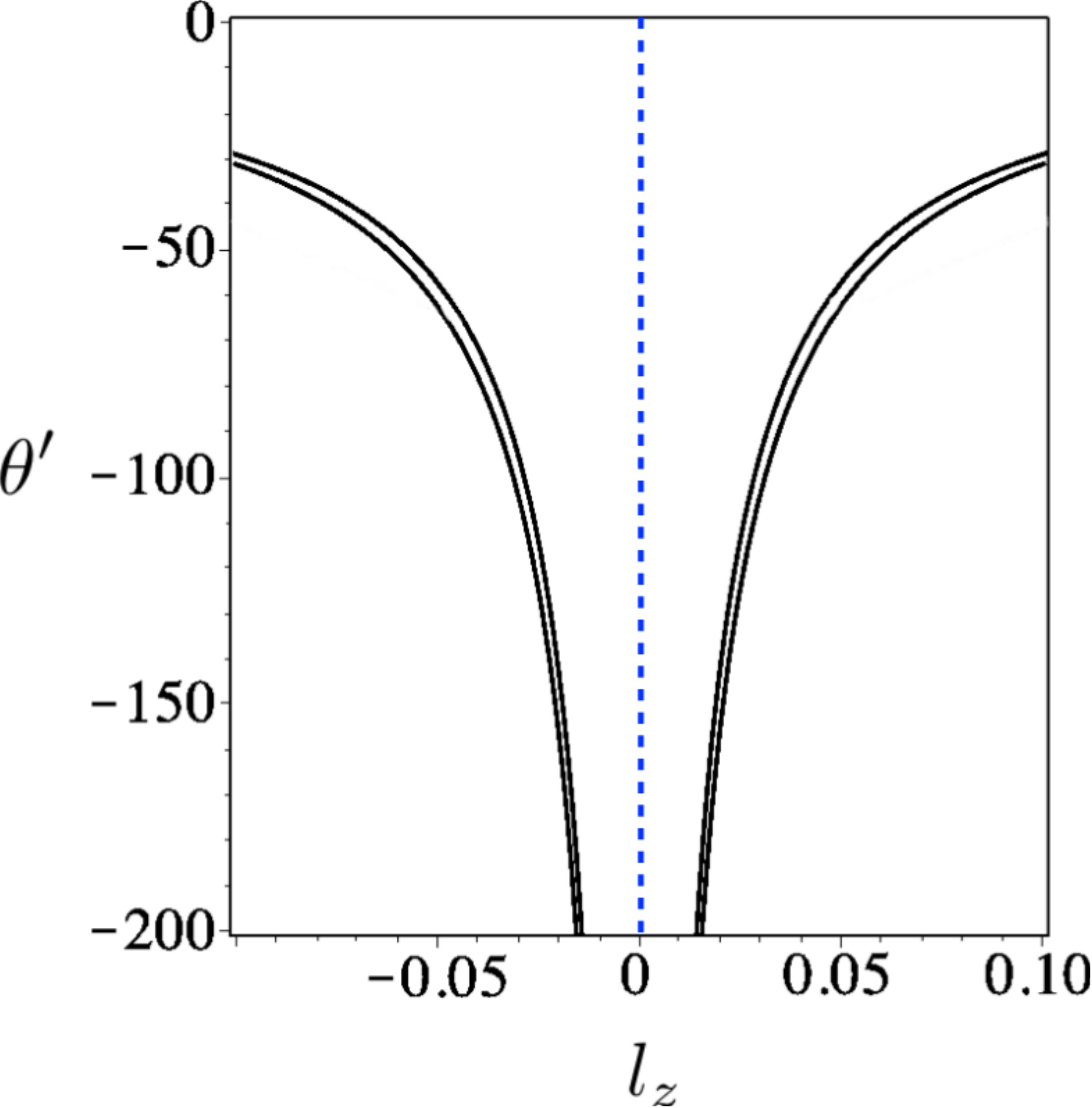}\hspace{0.5cm} &
	  \includegraphics[width=5 cm]{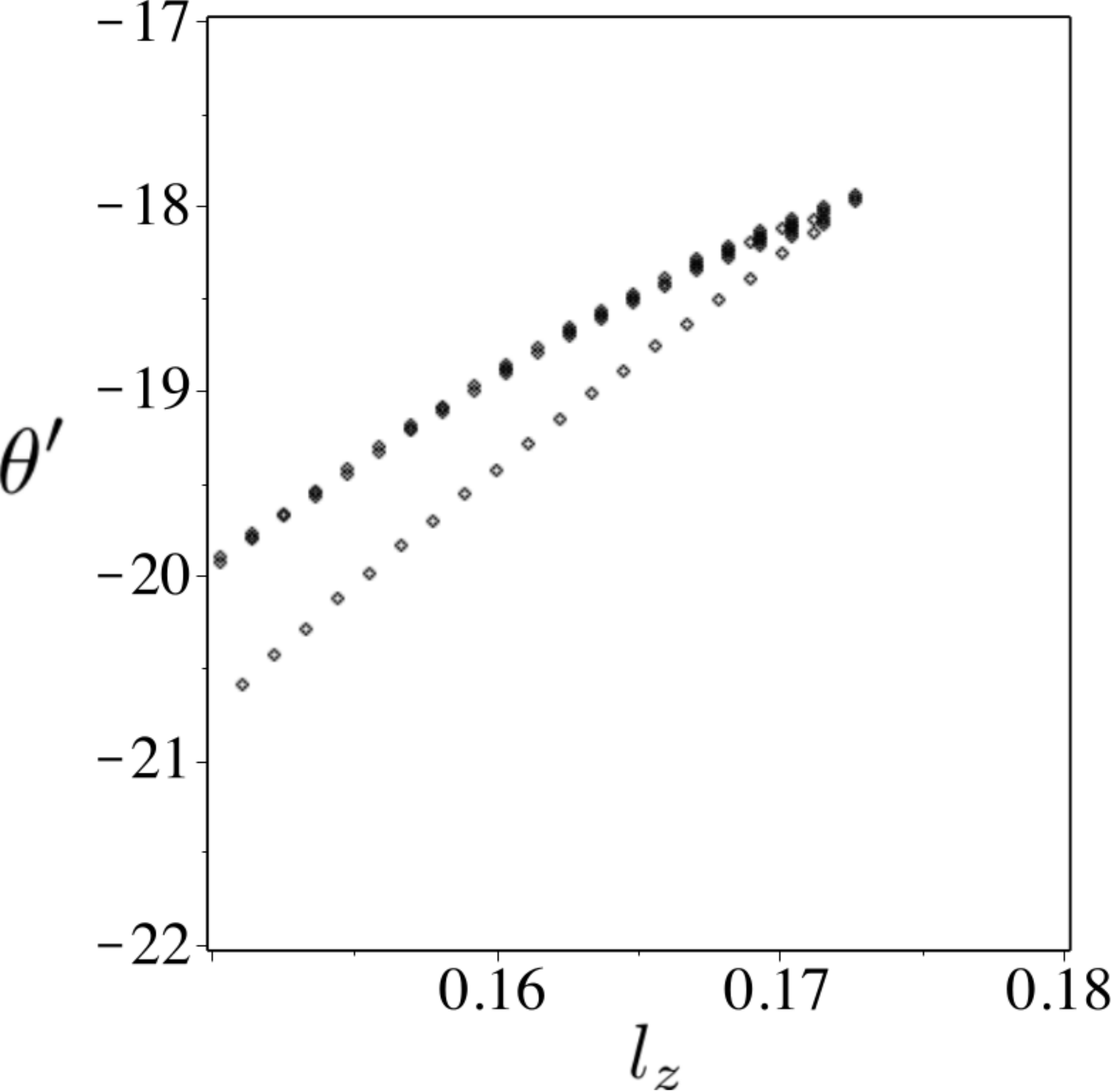} \\
  \end{tabular}}
         \caption{Value of ${\theta'}_{o}$ and angular momentum $l_z$ for an observer located on the equator and $r_o=5$, for $c_3=1/800$; on the left: the points with value of ${\theta'}_{o}$ and $l_z$ in the area denoted by S belong to the primary large shadow, in the middle:  the points with value of ${\theta'}_{o}$ and $l_z$ between the two solid lines belong to the secondary eyebrow shadow, in the right: zooming of the corner of the middle plot, for the range of $l_z=[0.15,0.18]$ and ${\theta'}_{o}=[-17,-22]$. We see that the two solid lines merge together. }\label{f7}
\end{figure}

As we mentioned before the values of $p$ and $q$ of a trajectory are related to ${\theta'}_{o}$ and angular momentum, ($l_z$), at the point $r_o=5$ and $\theta=\pi/2$. In figure \ref{f7}, we illustrate the corresponding values of ${\theta'}_{o}$ and $l_z$ for the points ($p, q$), given in Figures \ref{f3} and \ref{f5} for $c_3=1/800$. Points with values of ${\theta'}_{o}$ and $l_z$ in the area denoted by $S$ belong to the primary image of the black hole shadow. The region denoted by $S$ is closed as ${\theta'}_{o}\rightarrow + \infty$ and $l_z \rightarrow 0$, which is not possible to  illustrate in the figure. For the secondary image of the shadow the same thing holds but now
for ${\theta'}_{o}\rightarrow - \infty$ and $l_z \rightarrow 0$. In our map $(p,q)\rightarrow ({\theta'}_{o},l_z)$ we cannot consider $q$ very small (corresponds to $l_z \rightarrow 0$) whilst retaining finite-sized $p$. 
On the right side of Figure \ref{f7}, we can see how the region belonging to the eyebrow closes for the upper-left and upper right corners. From this figure, and from the trajectory graphs (discussed later), for $l_z, {\theta'}_{o}$ both very small the photon escapes, in contrast to the undistorted case. 

In the Appendix, we use the analogy with the Newtonian picture in order to give an interpretation for multiple moments. In the Newtonian case that, $\tilde{c}_3=-\tilde{c}_1 >0$, we can see that, \[\frac{m_1}{{r_1}^2}>\frac{m_2}{{r_2}^2} \, , \] and the gravitational field of the upper source is stronger. In our case, when $c_3>0$ more photons with positive velocities ${\theta'}_{o}>0$ are getting absorbed. Therefore, they create a bigger shadow as we see in the left side of figure \ref{f7}; this effect would appear to be due to the   the gravitational field being stronger on the upper plane than the lower one. However, in the $(p,q)$-plane we see that  the primary image of the shadow is in the lower plane  (example see left side of Figure \ref{f3}). This is because we made the choice of ${\theta'}_{o}\rightarrow - p$. One could make a different choice such that ${\theta'}_{o}\rightarrow  p$, which is equally correct.
\begin{figure}[htp]
\begin{center}
\setlength{\tabcolsep}{ 0 pt }{\scriptsize\tt
		\begin{tabular}{ cc }
		\includegraphics[width=5 cm]{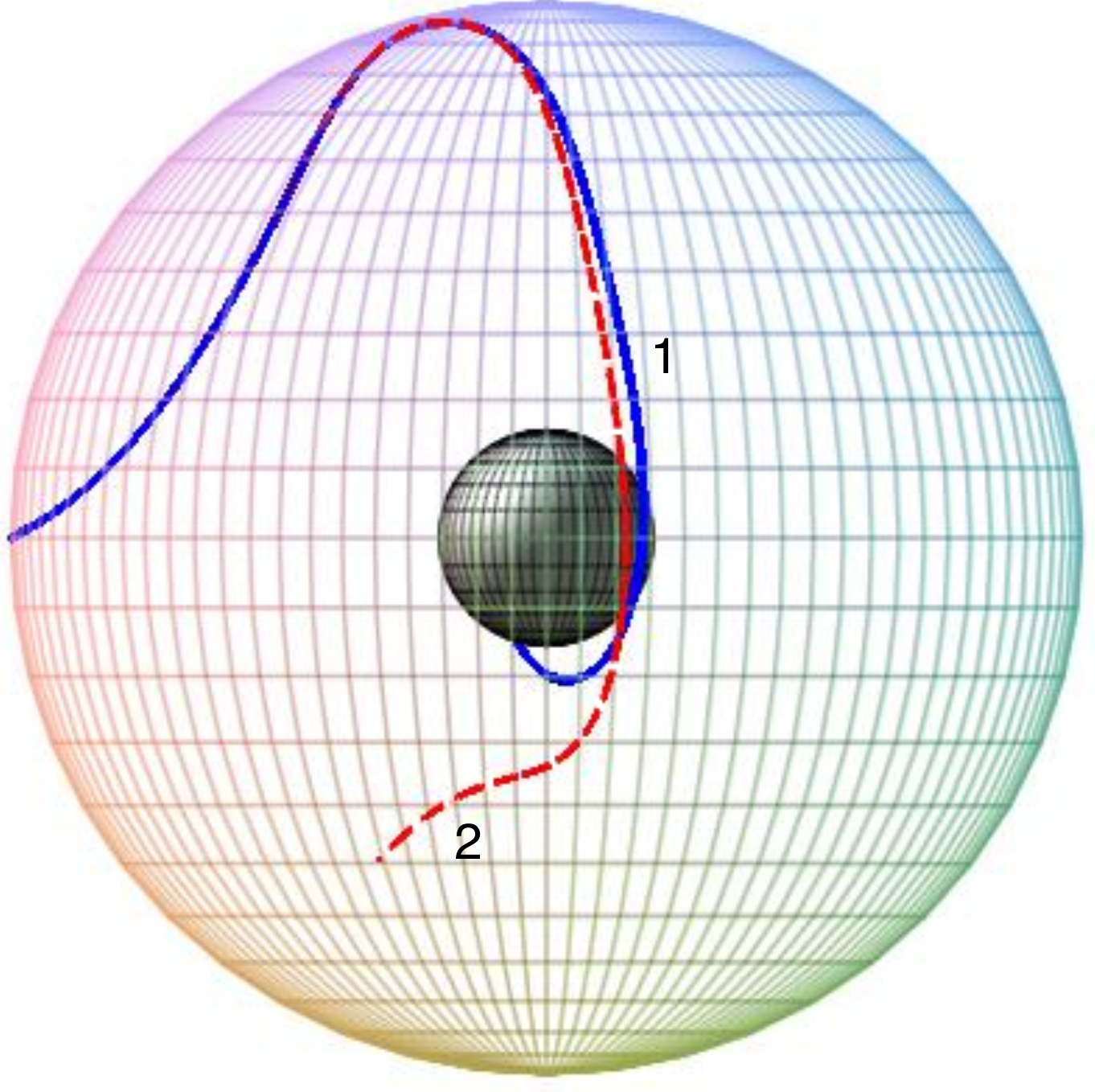}&
		 \hspace{2cm} \includegraphics[width=5 cm]{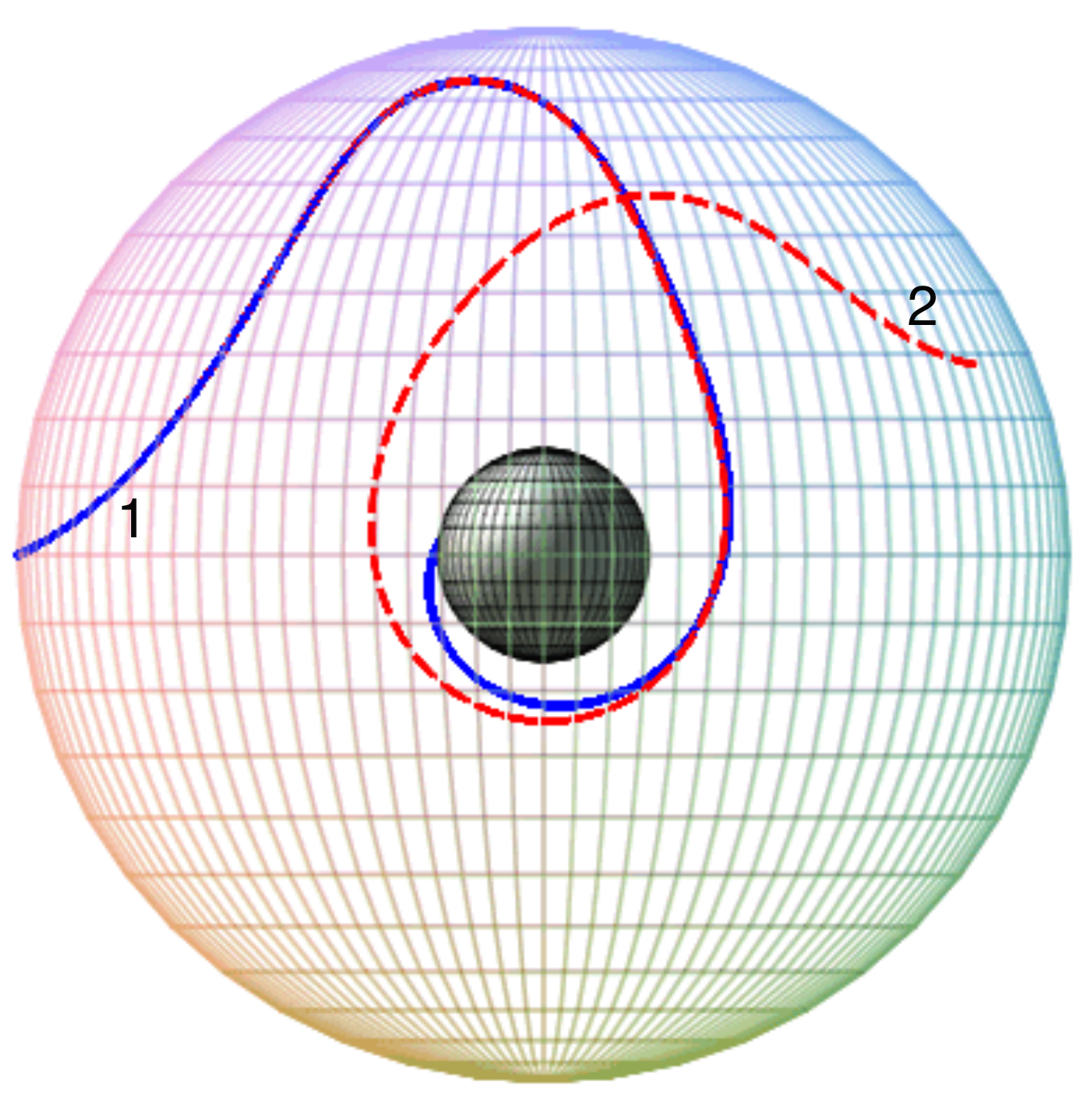} \\
		   \end{tabular}}
		   \end{center}
        \caption{Depiction of photon trajectories for  $c_3=1/800$: the starting points of the trajectories are where the photon enters  the sphere of radius $r_o=5$; the endpoints are where photon  either leaves the sphere or enters the black hole. On the left side: Trajectories of a photon for $p=2.02$ and $q=0.14$, ( $l_z=-0.157$, ${\theta'}_{o}=-19.62$) in blue-solid line (line 1) and for $p=2.02$ and $q=0.17$, ($l_z=-0.19$, ${\theta'}_{o}=-16.16$) in red-dashed line (line 2). On the right side: Trajectories of a photon for $p=1.89$ and $q=0.04$, ($l_z=-0.04$, ${\theta'}_{o}=-64.27$) in blue-solid line (line 1) and  for $p=1.89$ and $q=0.07$, ($l_z=-0.08$, ${\theta'}_{o}=-36.72$)in red-dashed line (line 2). At the beginning the lines are joint. The distortions of the black hole located at the centre of the sphere are too small to illustrate.}\label{f8}       
\end{figure} 
\begin{figure}[htp]
\begin{center}
\setlength{\tabcolsep}{ 0 pt }{\scriptsize\tt
		\begin{tabular}{ cc }
		\includegraphics[width=5 cm]{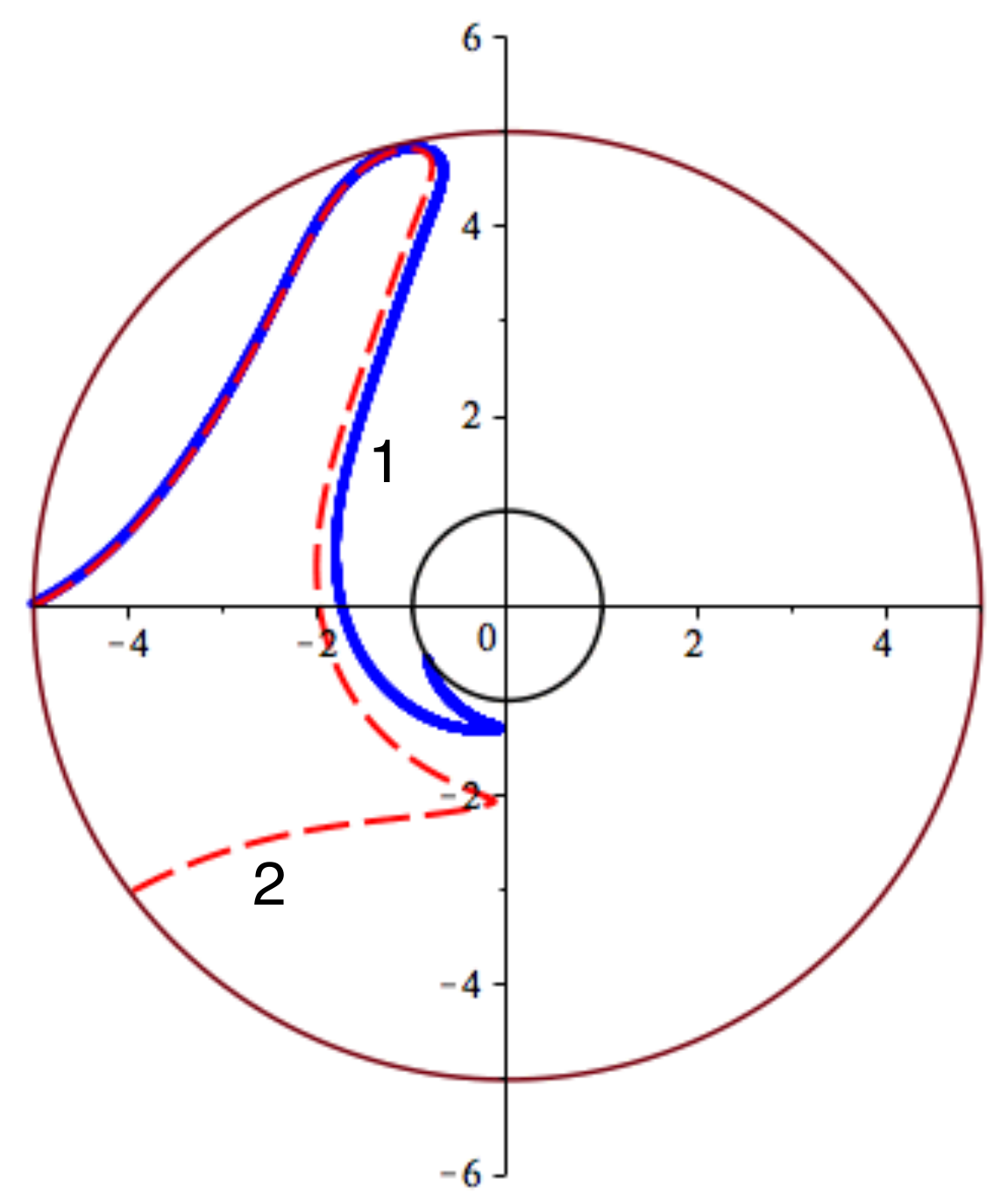}&
		 \hspace{2cm} \includegraphics[width=5 cm]{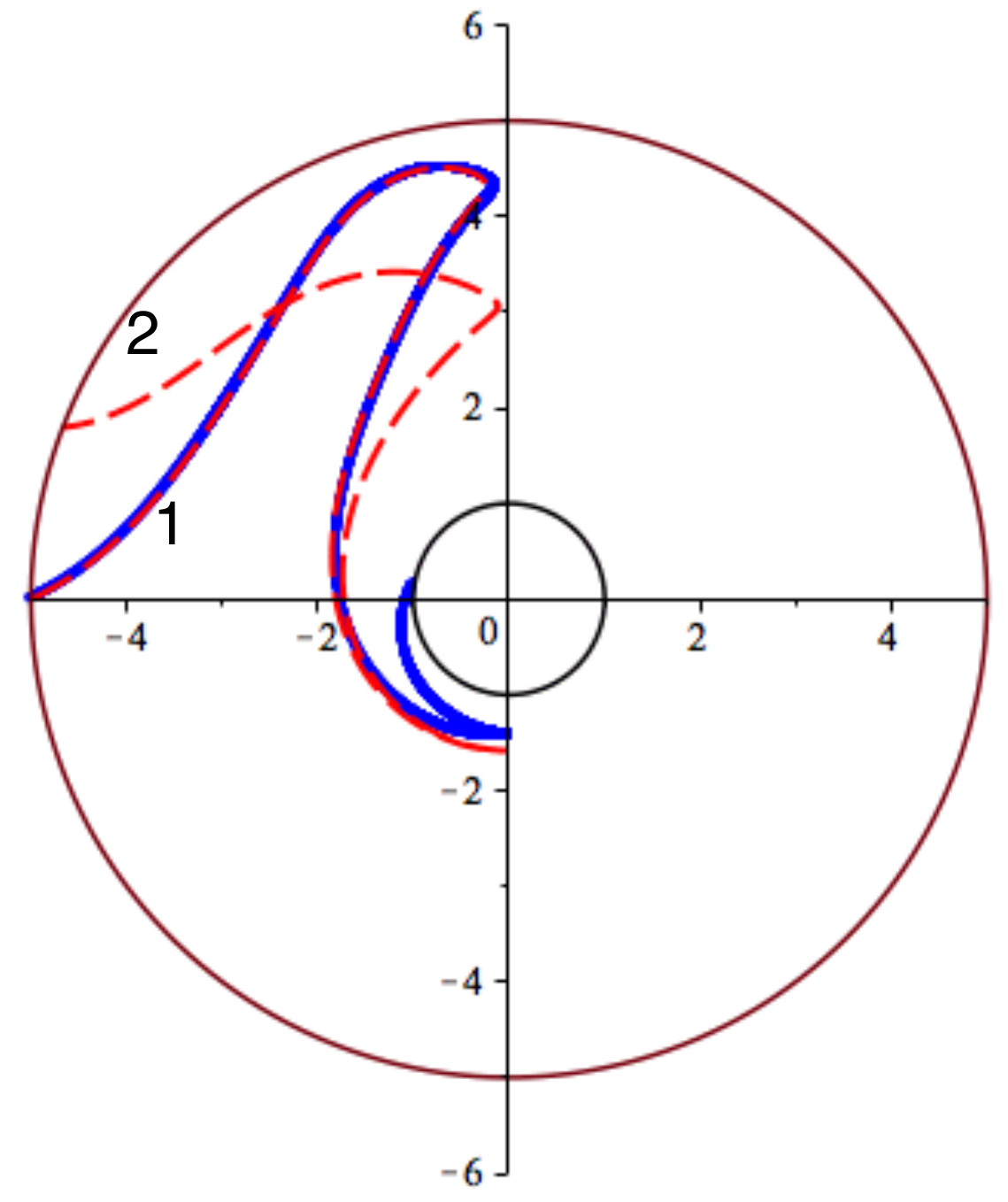} \\
		 $~p=2.02$ and $q=0.14$ in blue-solid (line 1), & $~p=1.89$ and $q=0.04$ in blue-solid (line 1),\\
		 $p=2.02$ and $q=0.17$ in red-dashed (line 2). & $p=1.89$ and $q=0.07$ in red-dashed (line 2).\\
		   \end{tabular}}
		     \end{center}
        \caption{Depiction of the  $r$- and $\theta$- coordinates of the trajectories in the Fig. \ref{f8} projected at constant azimuthal angle $\phi$.}
        \label{f8-2D}       
\end{figure} 

\begin{figure}[htp]
\setlength{\tabcolsep}{ 0 pt }{\scriptsize\tt
		\begin{tabular}{ c }
	 \hspace{4.5cm}	\includegraphics[width=5 cm]{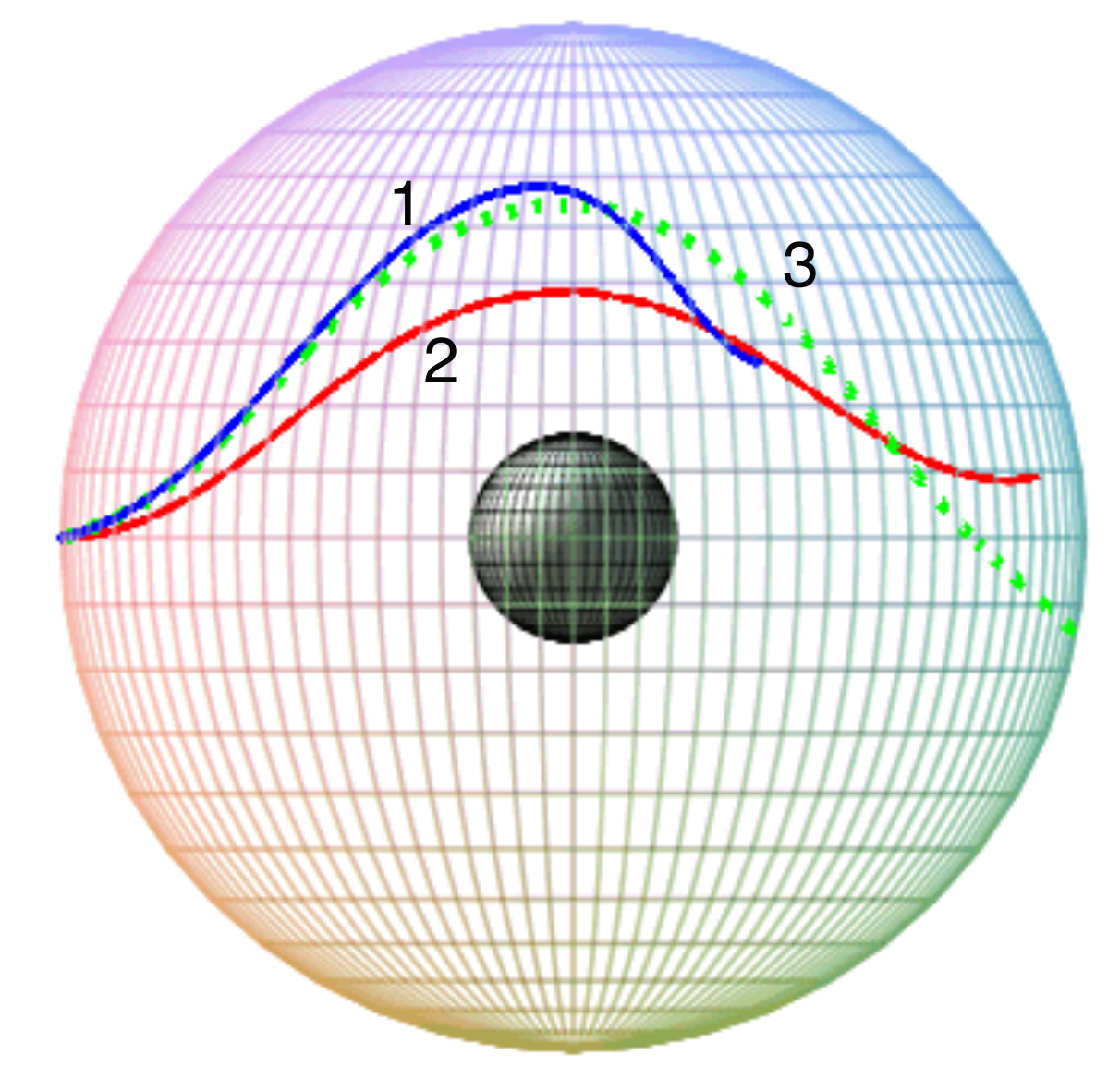}
	   \end{tabular}}
         \caption{Comparison of photon trajectories for  $c_3=1/800$ for differing values of $p$ and $q$, with start/endpoints
        meaning the same as in figure \ref{f8}. In blue-solid (line 1): trajectories of a photon for $p=1$ and $q=1$, ($l_z=-1.12$, and ${\theta'}_{o}=-1.36$). In red-dashed (line 2): for $p=0$ and $q=1$, ($l_z=-1.12$, and ${\theta'}_{o}=0$). In green-dotted (line 3): for $p=1$ and $q=0.001$, ($l_z=-0.001$, and ${\theta'}_{o}=-1360.1$)}\label{f10}
\end{figure}

\begin{figure}[htp]
\begin{center}
\setlength{\tabcolsep}{ 0 pt }{\scriptsize\tt
		\begin{tabular}{ cc c }
		\includegraphics[width=4.7 cm]{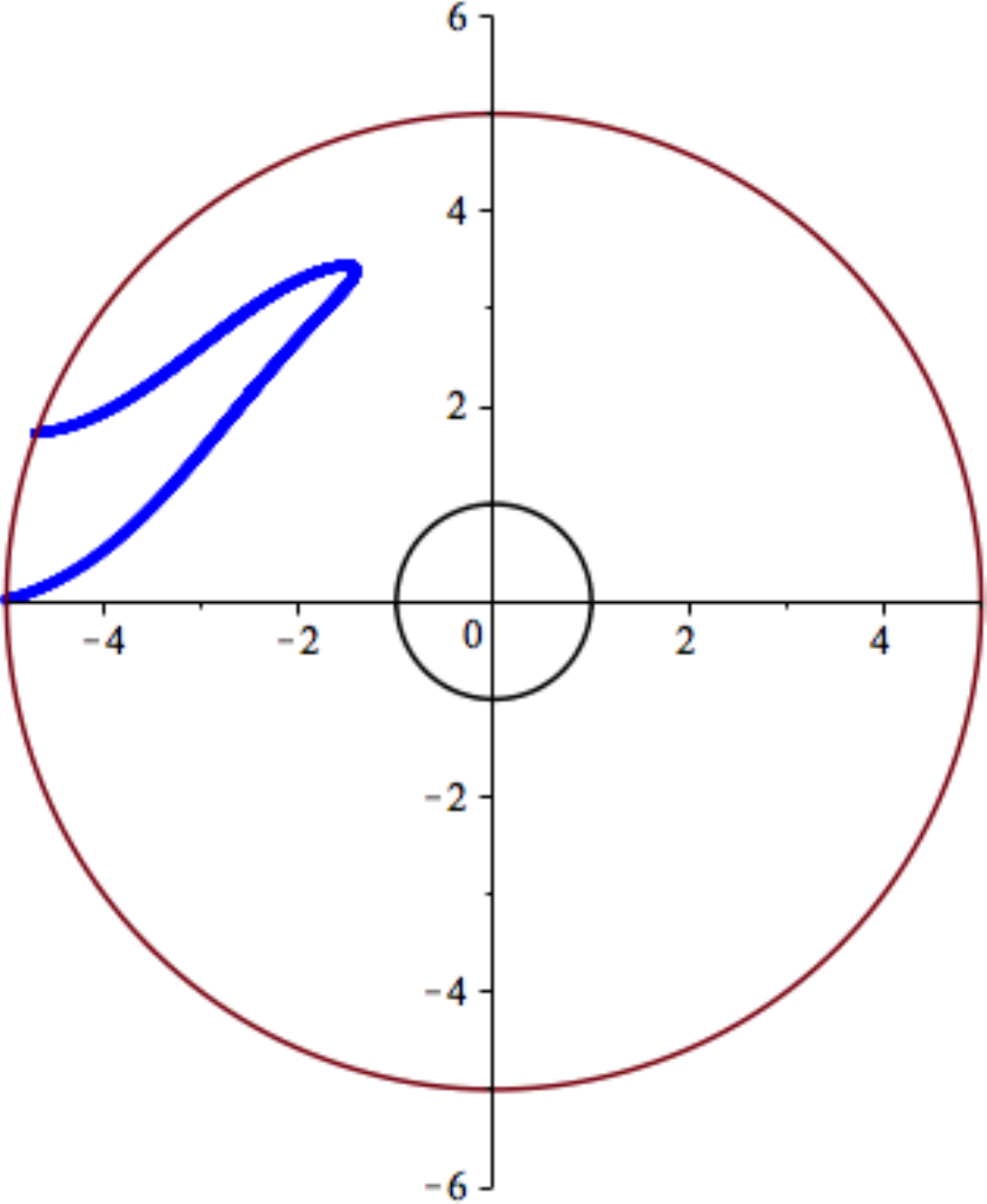} \hspace{0.5cm} &
	 \includegraphics[width=4.7 cm]{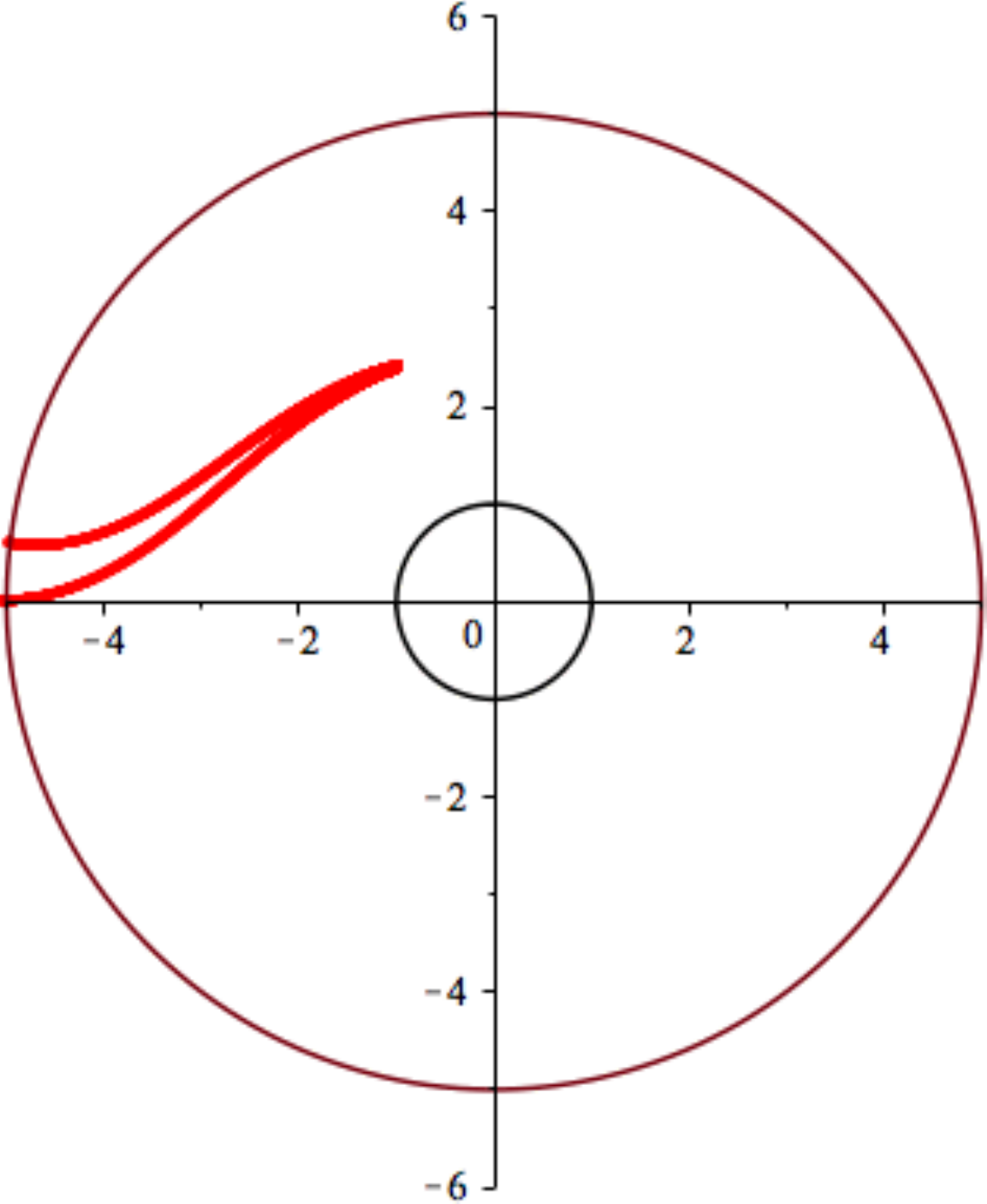}\hspace{0.5cm} &
	  \includegraphics[width=4.7 cm]{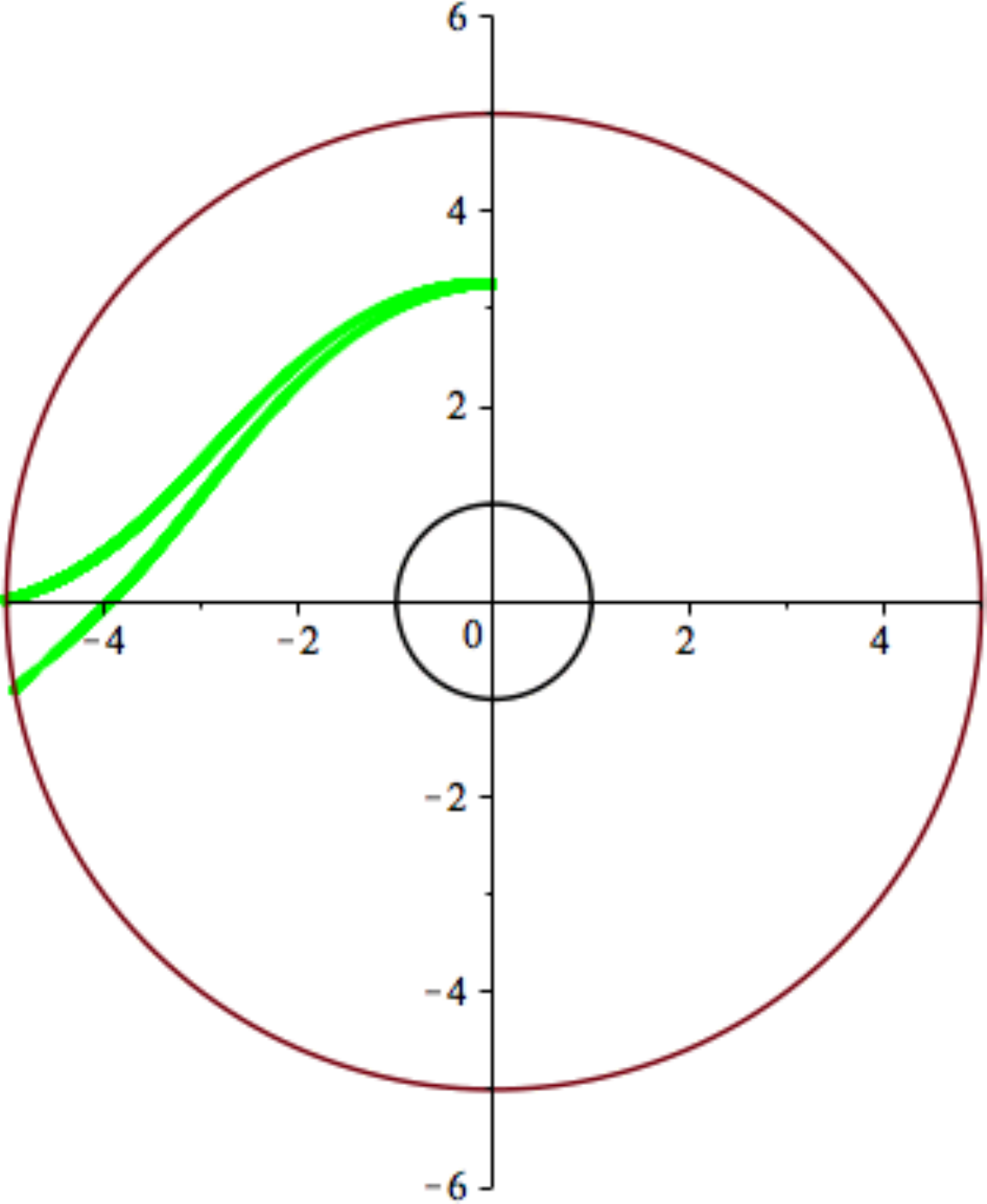} \\
	  $p=1$ and $q=1$ & $p=0$ and $q=1$ & $p=1$ and $q=0.001$. 
  \end{tabular}}
  \end{center}
         \caption{Depiction of the  $r$- and $\theta$- coordinates of the trajectories in   Fig. \ref{f10} projected at constant azimuthal angle $\phi$}\label{f10-2D}
\end{figure}

\begin{figure}[htp]
\begin{center}
\setlength{\tabcolsep}{ 0 pt }{\scriptsize\tt
		\begin{tabular}{ cc }
	\includegraphics[width=5 cm]{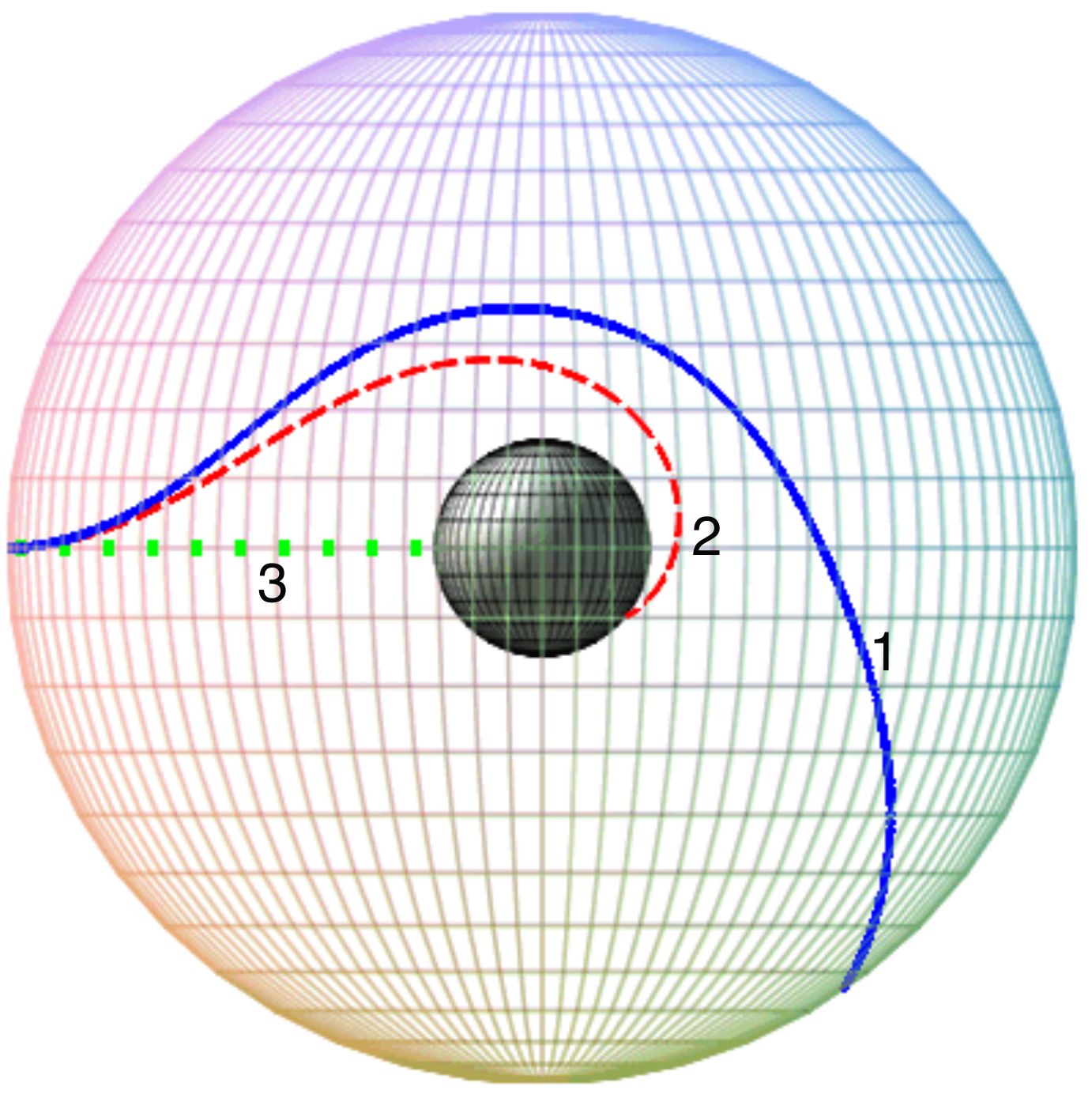}&
		 \hspace{2cm} \includegraphics[width=5 cm]{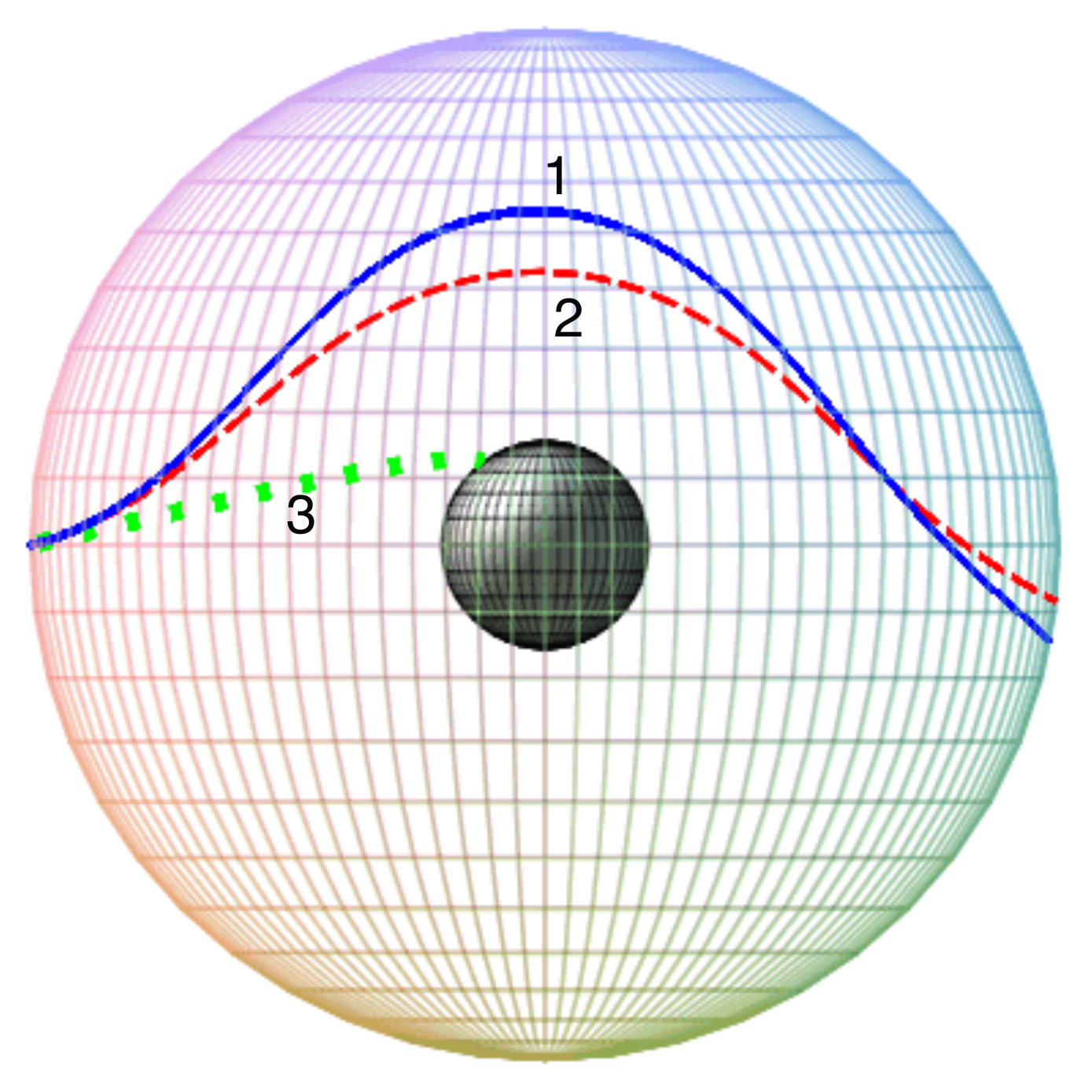} \\
		   \end{tabular}}
		   \end{center}
         \caption{Comparison of photon trajectories for three differing values of $c_3$, with  start/endpoints meaning the same as in figure \ref{f8}. On the left side: Trajectories of a photon for $p=0$, $q=0.001$, ($l_z=-0.001$, and ${\theta'}_{o}=0$) and $c_3=1/800$ in blue-solid (line 1), for $c_3=1/1000$ in red-dashed (line 2) and for $c_3=0$ in green-dotted (line 3). On the right side: Trajectories of a photon for $p=1$, $q=0.001$, ($l_z=-0.001$, and ${\theta'}_{o}=-1360.11$) and $c_3=1/800$ in blue-solid (line 1), for $c_3=1/1000$ ($l_z=-0.001$, and ${\theta'}_{o}=-1217.55$) in red-dashed (line 2) and for $c_3=0$ ($l_z=-0.001$ ${\theta'}_{o}=-1000$) in green-dotted (line 3).}\label{f9}
\end{figure}

\begin{figure}[htp]
\begin{center}
\setlength{\tabcolsep}{ 0 pt }{\scriptsize\tt
		\begin{tabular}{ cc c }
		\includegraphics[width=4.7cm]{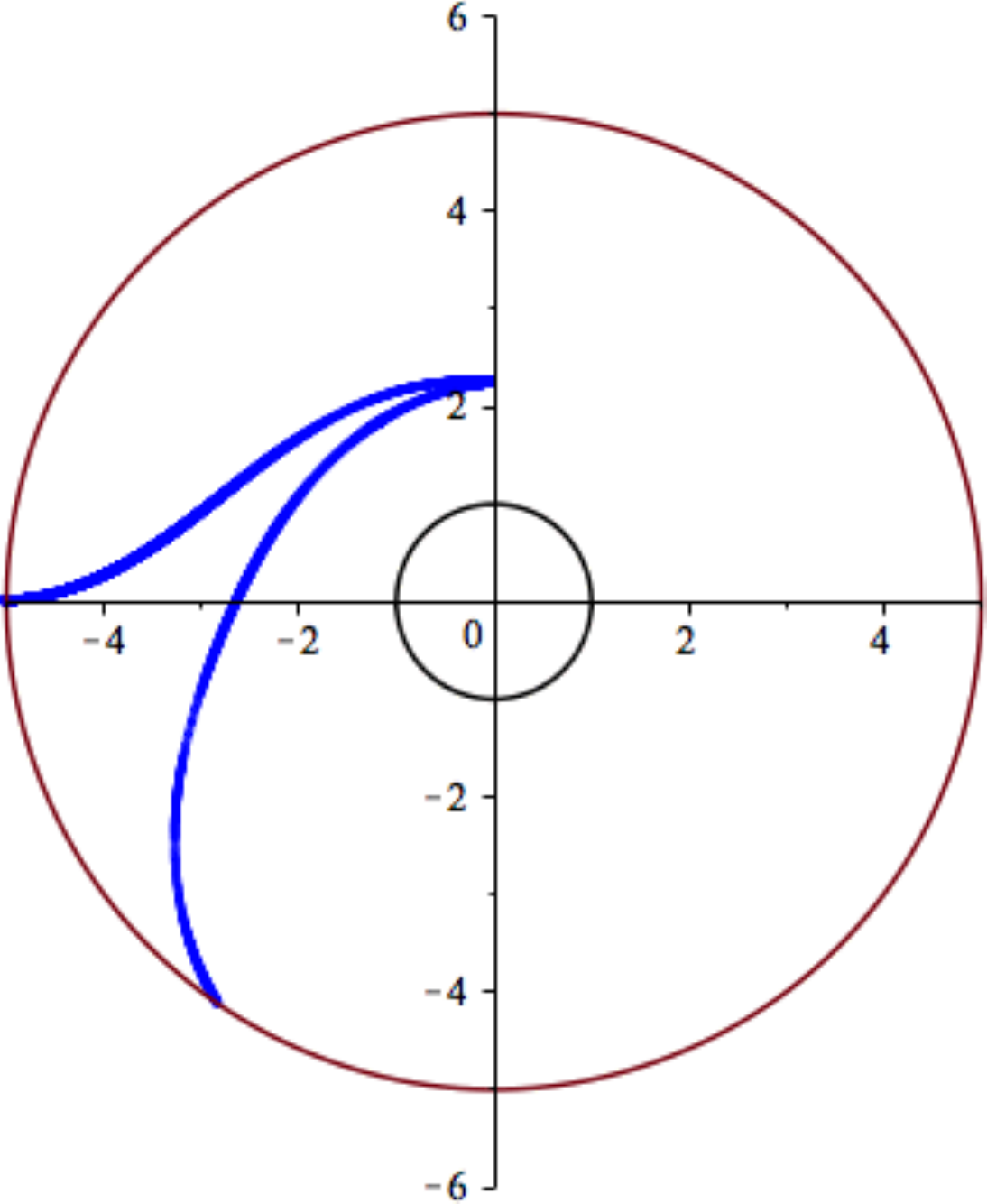} \hspace{0.5cm} &
	 \includegraphics[width=4.7cm]{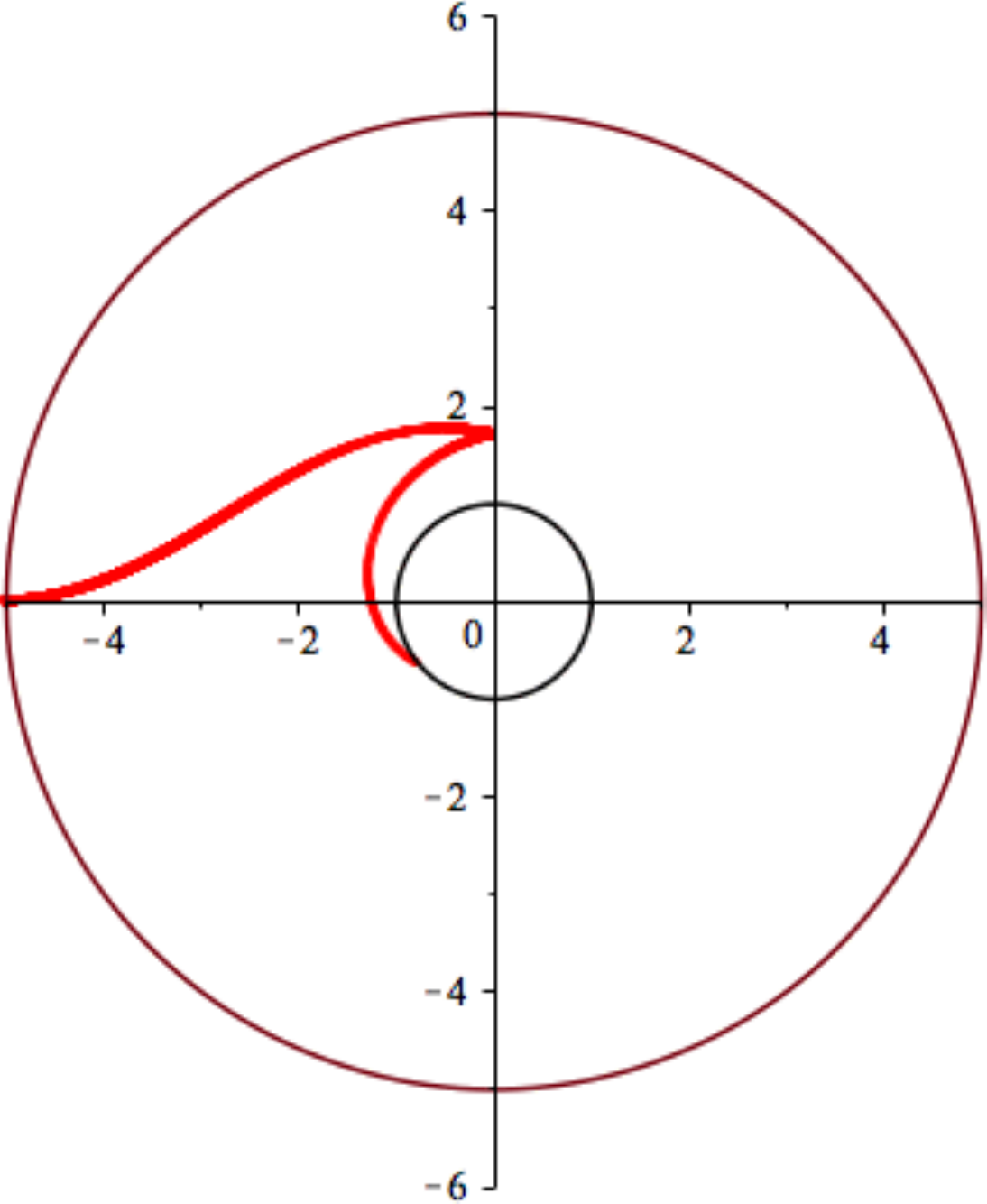}\hspace{0.5cm} &
	  \includegraphics[width=4.7cm]{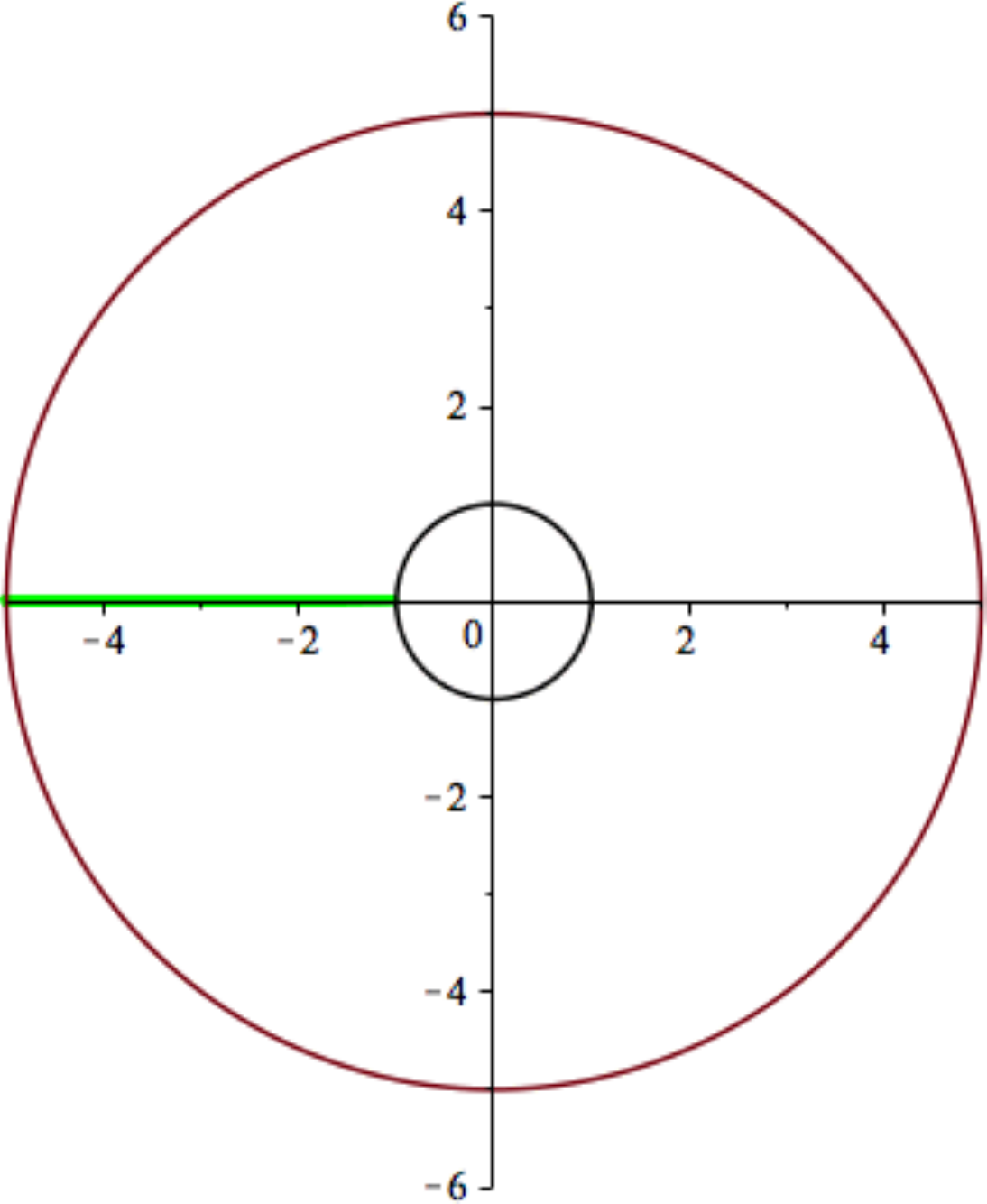} \\
	 $p=0$, $q=0.001$, and $c_3=1/800$ & $p=0$, $q=0.001$, and $c_3=1/1000$ &
	 $p=0$, $q=0.001$, and $c_3=0$ 
  \end{tabular}}
 \end{center}
         \caption{Depiction of the  $r$- and $\theta$- coordinates of the trajectories in the  left graph of  Fig. \ref{f9}, projected at constant azimuthal angle $\phi$}\label{f9-2D1}
\end{figure}

\begin{figure}[htp]
\setlength{\tabcolsep}{ 0 pt }{\scriptsize\tt
		\begin{tabular}{ cc c }
		\includegraphics[width=4.7 cm]{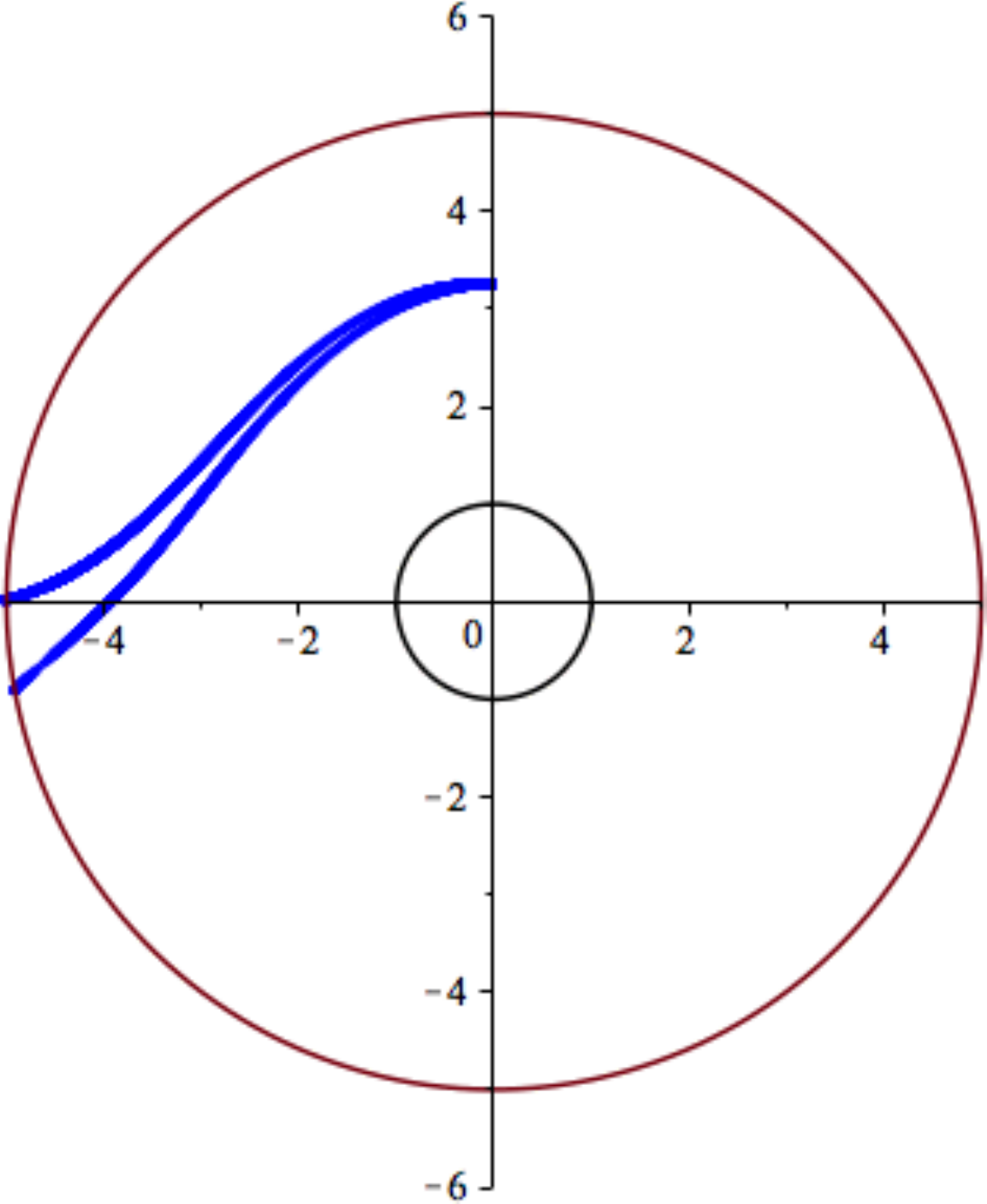} \hspace{0.5cm} &
	 \includegraphics[width=4.7 cm]{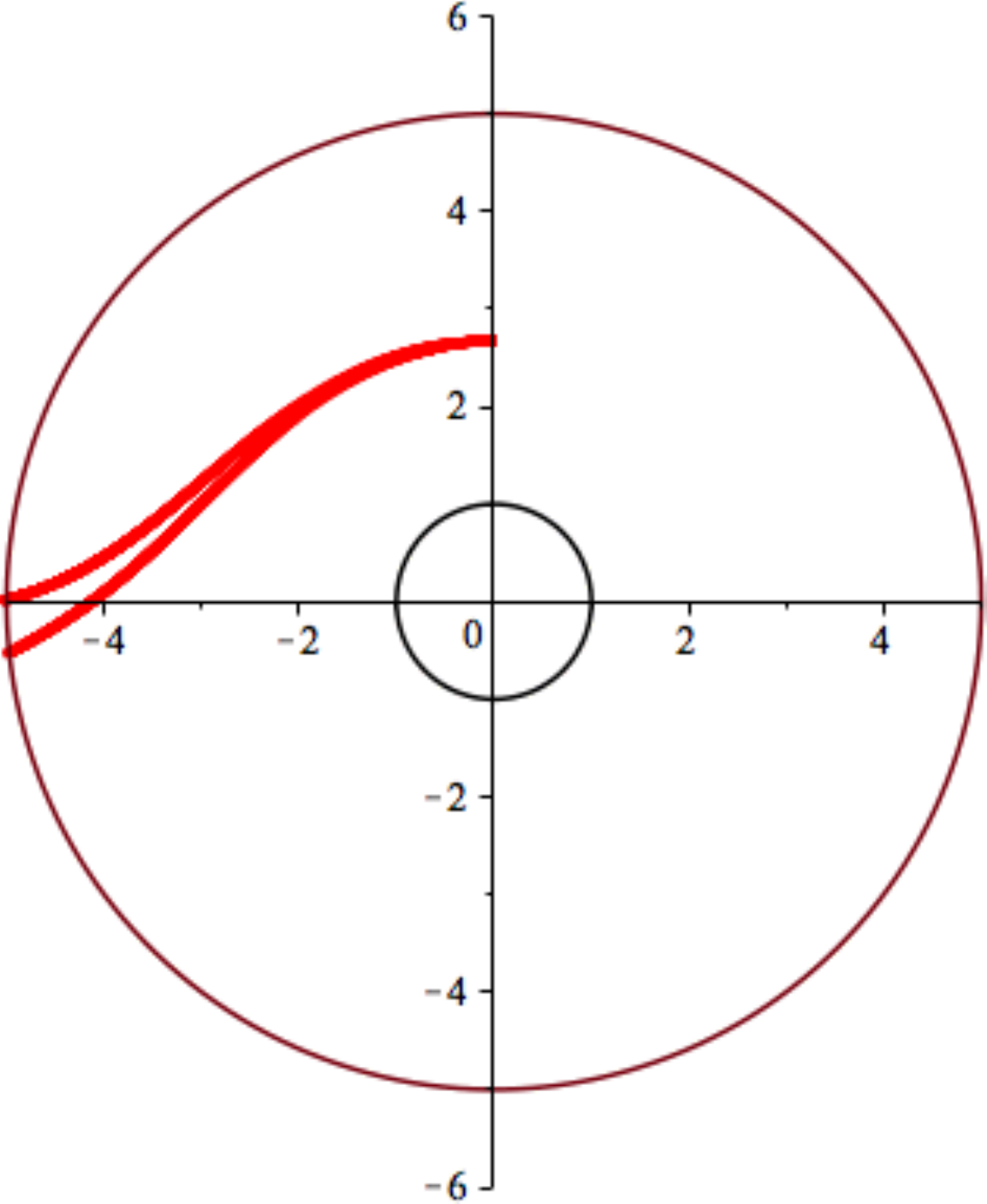}\hspace{0.5cm} &
	  \includegraphics[width=4.7 cm]{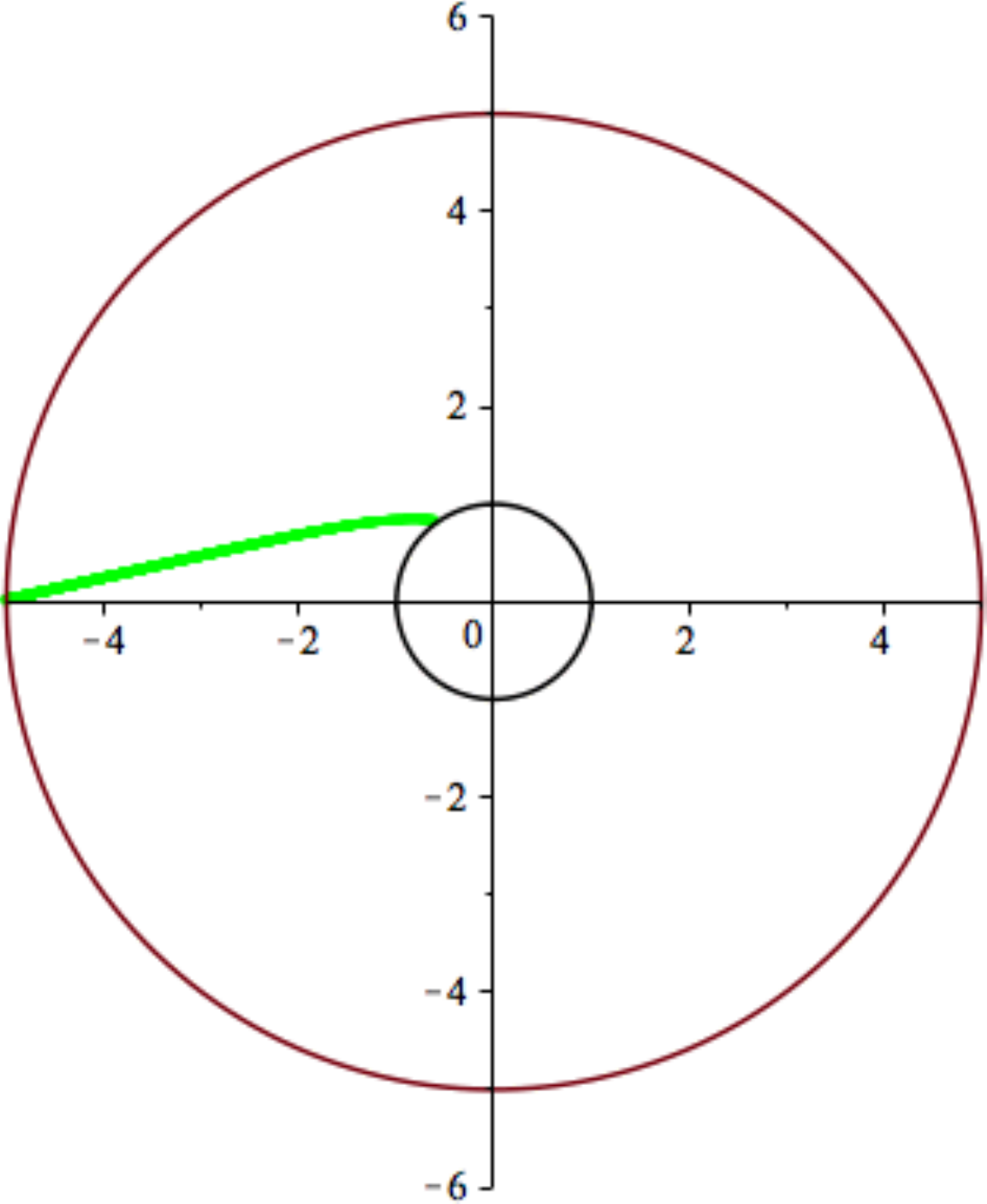} \\
	  $p=1$, $q=0.001$, and $c_3=1/800$ & $p=1$, $q=0.001$, and $c_3=1/1000$ &
	 $p=1$, $q=0.001$, and $c_3=0$
  \end{tabular}}
         \caption{Depiction of the  $r$- and $\theta$- coordinates of the trajectories in the right graph of  Fig. \ref{f9}, projected at constant azimuthal angle $\phi$}\label{f9-2D2}
\end{figure}

\begin{figure}[htp]
\setlength{\tabcolsep}{ 0 pt }{\scriptsize\tt
		\begin{tabular}{ cc c}
		\includegraphics[width=4.4 cm]{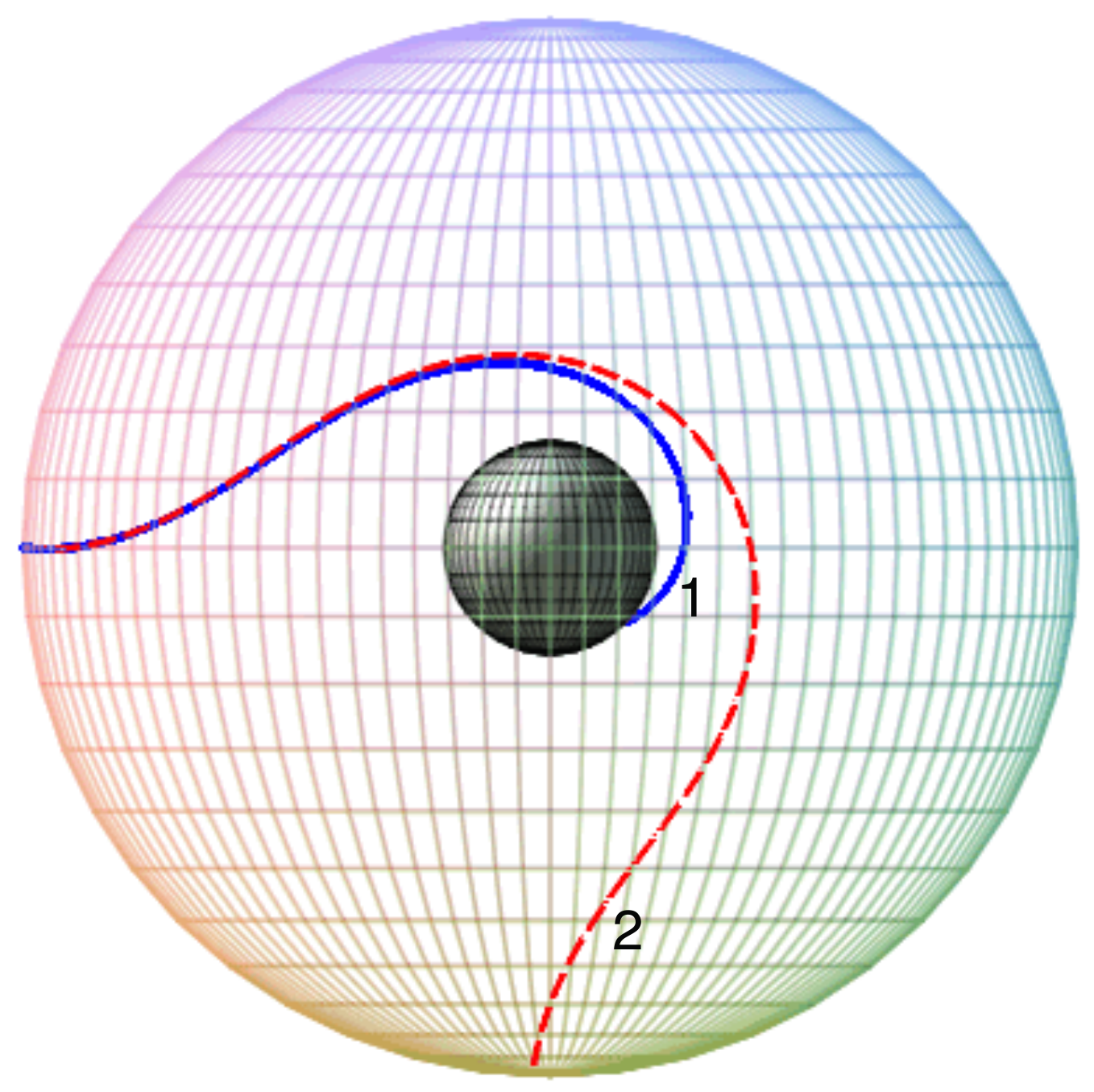}&
		 \hspace{1cm} \includegraphics[width=4.7 cm]{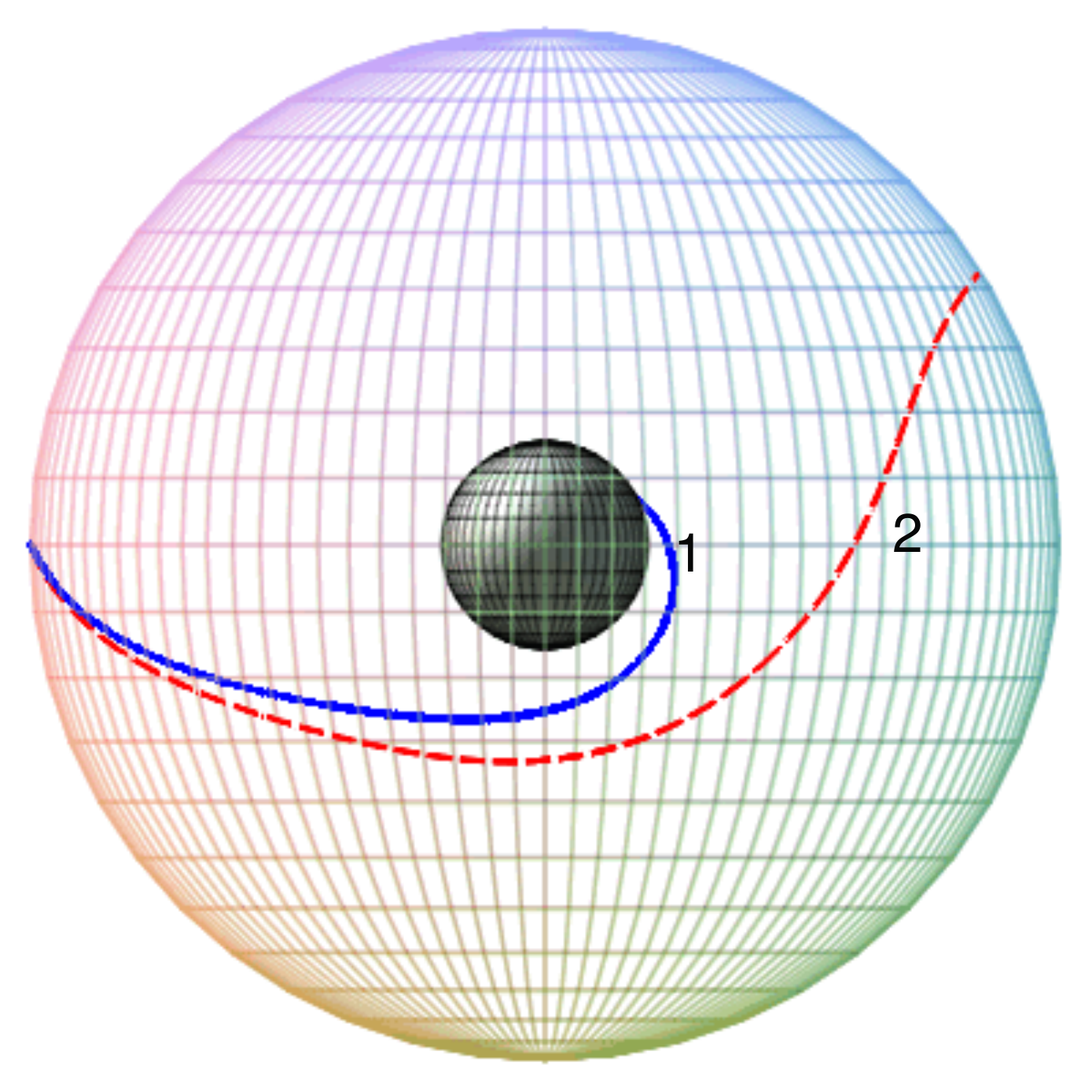}
		 \hspace{1cm} \includegraphics[width=4.7 cm]{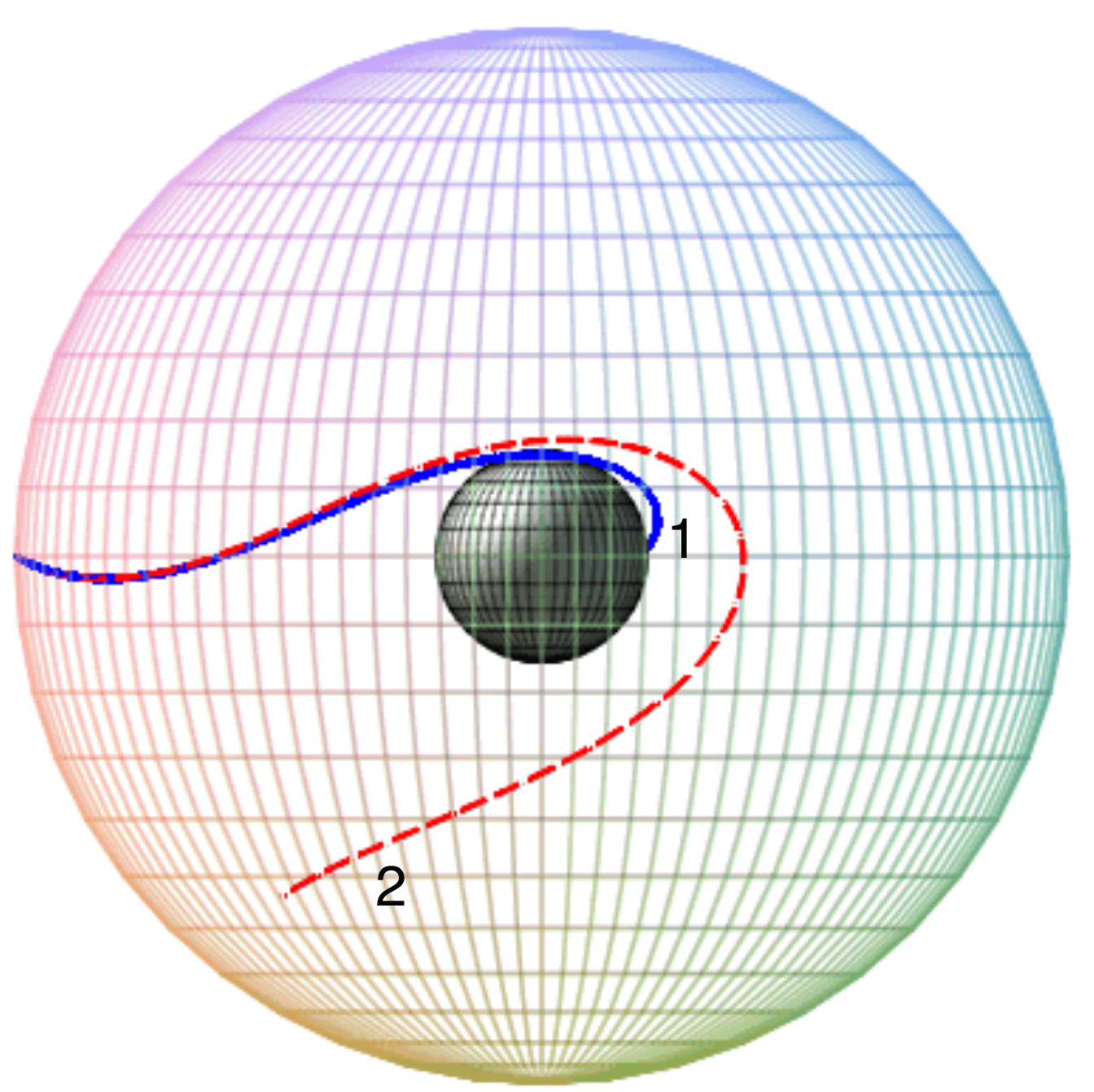} \\
		   \end{tabular}}
         \caption{Comparison of photon trajectories for  $cf_3=1/800$ for differing values of $p$ and $q$, with  start/endpoints meaning the same as in figure \ref{f8}. On the left side: Trajectory of a photon with $p=-0.6$, $q=0.2$, ($l_z=-0.22$, and ${\theta'}_{o}=3.00$) in blue-solid (line 1), for $p=-0.6$, $q=0.8$, ($l_z=-0.89$, and ${\theta'}_{o}=0.75$) in red-dashed (line 2). In the middle: trajectory of a photon with $p=-4.4$, $q=0.8$, ($l_z=-0.89$, and ${\theta'}_{o}=5.50$) in blue-solid (line 1), and $q=-4.6$, $p=0.2$, ($l_z=-0.22$, and ${\theta'}_{o}=23.00$) in red-dashed (line 2). On the right: trajectory of a photon with $p=-2.2$, $q=1.7$, ($l_z=-1.9$, and ${\theta'}_{o}=1.29$) in blue-solid (line 1), for $p=-2.2$ and $q=1.9$, ($l_z=-2.12$, and ${\theta'}_{o}=1.16$) in red-dashed (line 2).}\label{f11}
\end{figure}

\begin{figure}[htp]
\setlength{\tabcolsep}{ 0 pt }{\scriptsize\tt
		\begin{tabular}{ cc c}
		\includegraphics[width=4.2 cm]{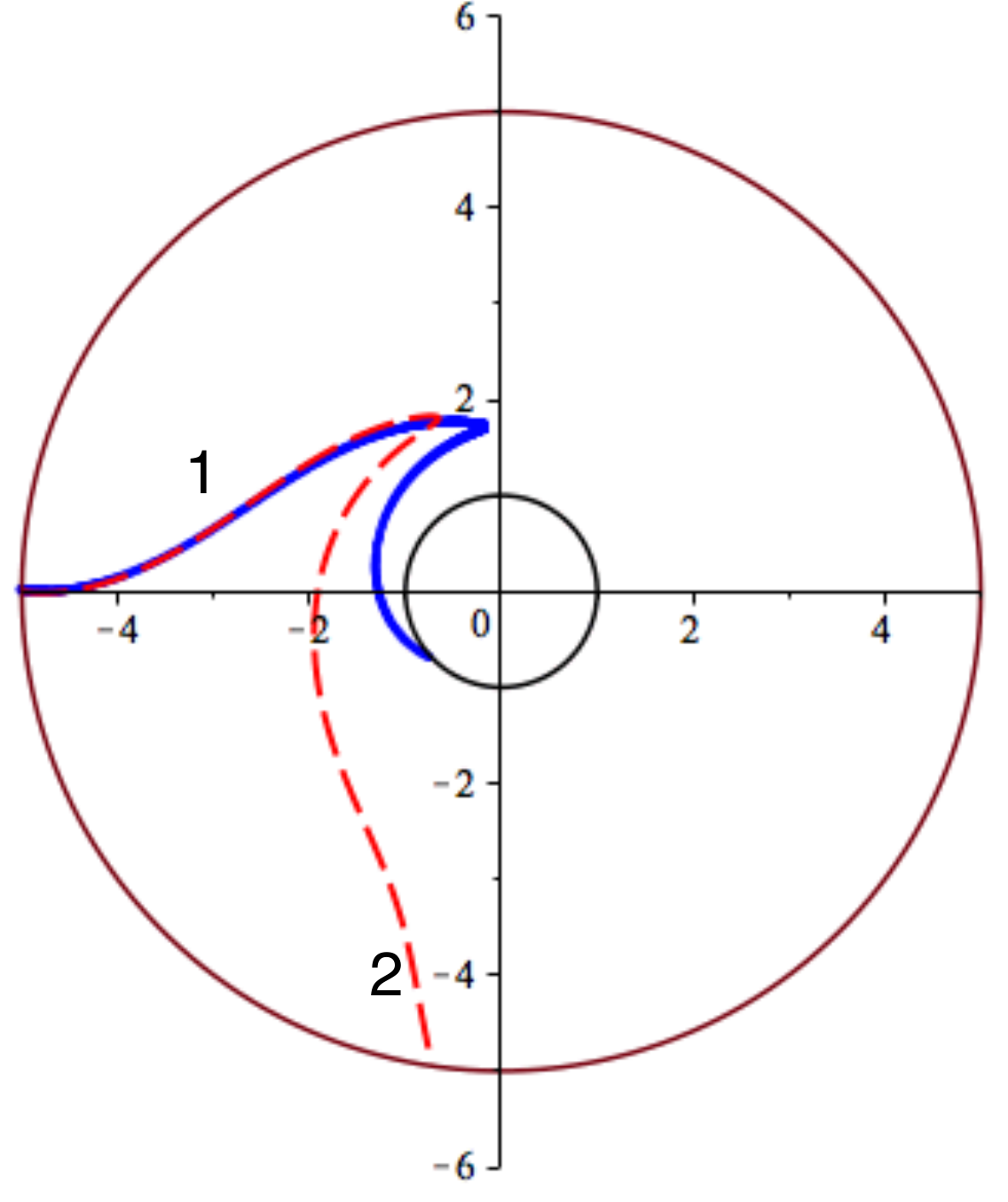}&
		 \hspace{1cm} \includegraphics[width=4.2 cm]{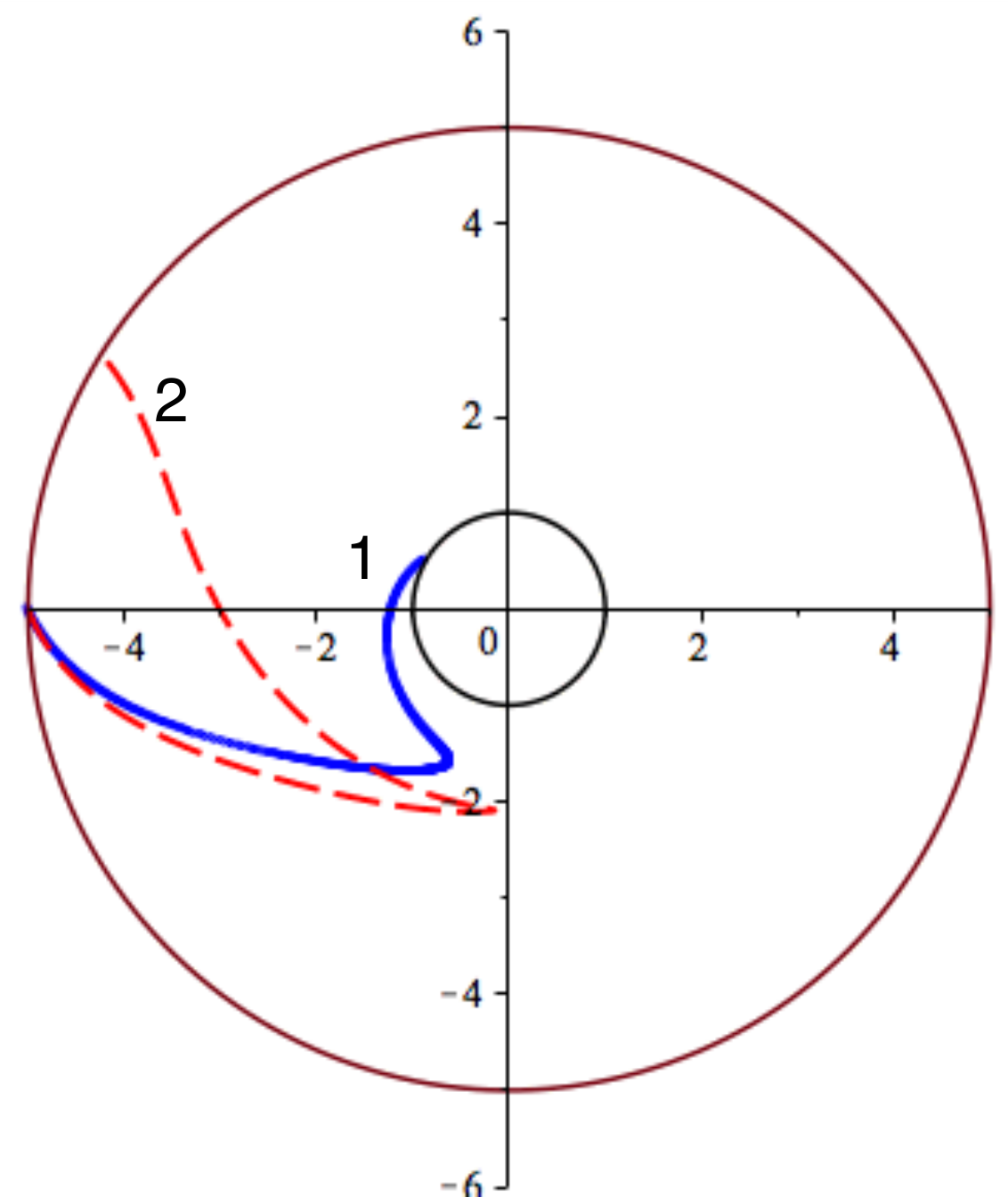}
		 \hspace{1cm} \includegraphics[width=4.2 cm]{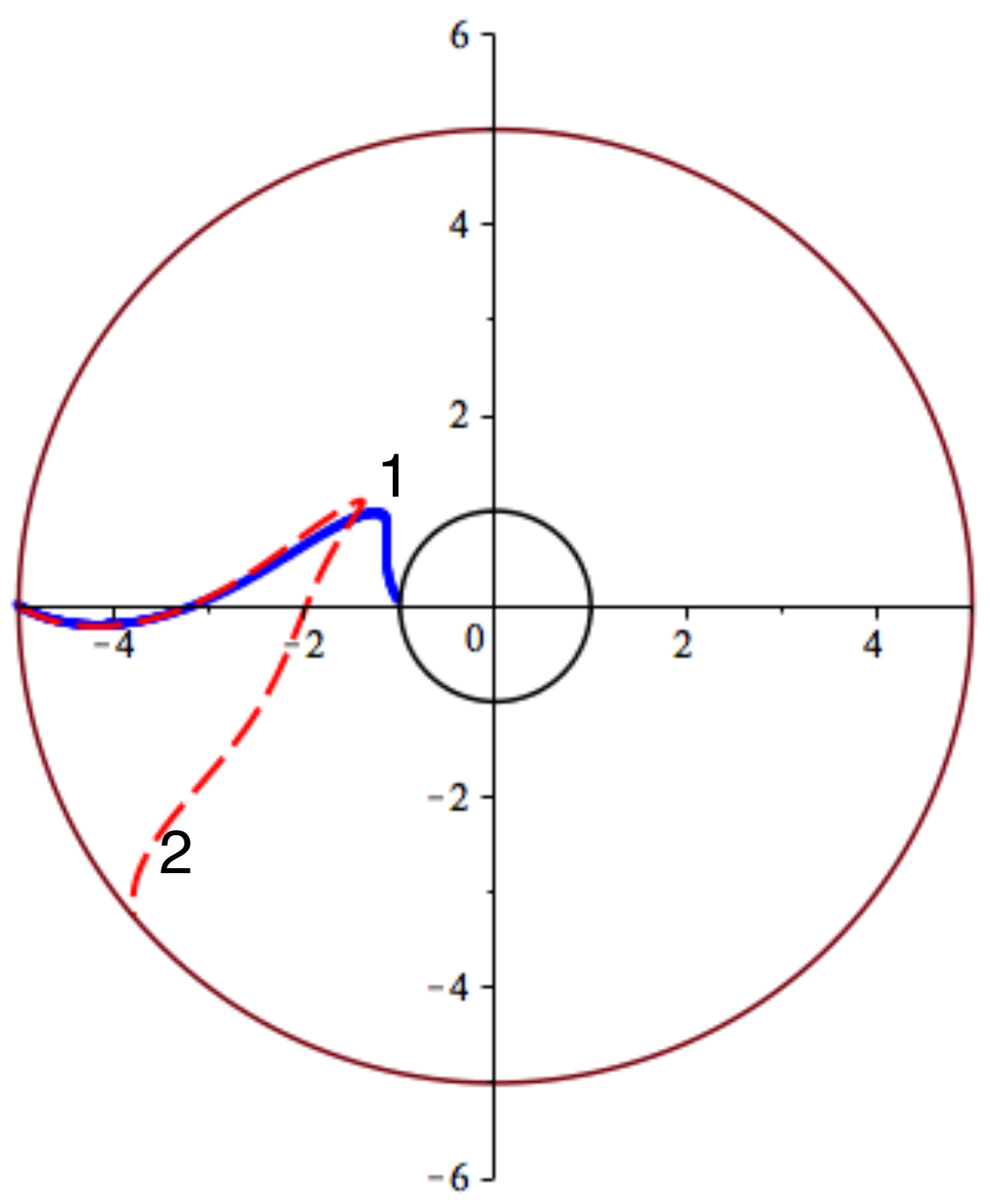} \\
		$~p=-0.6$, $q=0.2$ in blue& \hspace{-5cm}$~p=-4.4$, $q=0.8$ in blue & 
		\hspace{-4.5cm}$p=-2.2$, $q=1.7$ in blue \\
	$p=-0.6$, $q=0.8$ in red-dashed & \hspace{-5cm}$q=-4.6$, $p=0.2$ in red-dashed &\hspace{-4.3cm} $~p=-2.2$ and $q=1.9$ in red-dashed	
		   \end{tabular}}
         \caption{Depiction of the  $r$- and $\theta$- coordinates of the trajectories in the  left graph of  Fig. \ref{f11}, projected at constant azimuthal angle $\phi$.}\label{f11-2D}
\end{figure}

In figures \ref{f8}-\ref{f11-2D}, we illustrate the trajectories of a photon for given values of $p$ and $q$. We present each trajectory in two different figures, one including the $(r,\theta,\phi)$ coordinates of the trajectory (three-dimensional plots), the other including only $(r,\theta)$ coordinate of the trajectory (a two-dimensional plot projected at constant azimuthal angle). In the three-dimensional figures, the black sphere is the black hole horizon (at $r=1$) and the larger transparent sphere has radius $r_e=5$. In two-dimensional figures the red circle is at radius $r_e=5$ and the black circle corresponds to $r=1$ where the black hole horizon is located. The starting point of each trajectory is $r_o=5$ and $\theta_o=\pi/2$ (since we are tracing the trajectories backward, from the observation to the emission point). 

Our goal is not to categorize all the possible trajectories, but to compare some specific trajectories. For example 
in figure \ref{f11} we compare trajectories with values of $(p,q)$ that are inside the big primary shadow close to its rim with  trajectories having values of $(p,q)$ outside the big primary shadow close to its rim. The trajectories that have their endpoints on the transparent sphere do not belong to the shadow, whereas the ones that have their endpoints on the black sphere belong to the shadow. In figures \ref{f8} and \ref{f8-2D}, we illustrate the trajectories of a photon for distortion parameter $c_3=1/800$ and four different sets of initial conditions ($p,q$). The starting points of all these trajectories are at $r_{o}=5 ,\theta=\pi/2$, (since we are tracing the trajectories backward). These points either belong to the eyebrow structure or they are very close to the rim of it. In figure \ref{f10} and \ref{f10-2D}, we present the trajectory of a photon with three sets of initial conditions ($p,q$) between the eyebrow structure and the primary image of the black hole. In this figure, the distortion parameter remains $c_3=1/800$.

In figures \ref{f9}-\ref{f9-2D2}, we can see how the trajectory changes for the same initial conditions ($p,q$), whilst the distortion parameter takes the values $c_{3}=1/800$, $c_{3}=1/1000$ and $c_{3}=0$, (the undistorted case). In the absence of distortions a photon with $p=0$ and $q=0.001$ corresponding to a point very close to the centre of the ($p,q$)-plane comes very close to the horizon of the black hole (i.e., it is captured). However, in the case of distortion parameter $c_3\ne 0$ we can see how the trajectory with the same initial condition is modified, letting the photon to escape. On the left side, we have the point $p=0$ and $q=0.001$ and on the right side we have the point $p=1$, $q=0.001$. From the left side of figure \ref{f9}, we see that, while the distortion parameter decreases the shadow moves to the centre of ($p,q$)-plane. We can see this from the fact that, the same point for the case of $c_3=1/800$ is escape and for the cases of $c_3=1/1000$ and $c_3=0$ is capture. From the right side of Figure \ref{f9}, we can see that the same point is escape only for the case $c_3=0$.  In the Figures \ref{f11} and \ref{f11-2D}, we have six sets of initial conditions for the case that the distorted parameter is $c_3=1/800$. In each case, one set of initial conditions belongs to the primary shadow and we can see the trajectory of it before it gets captured and the second one is close to the rim of the primary shadow which escapes.   

In general we can say that in the case of $c_3>0$, the photons that travel from equatorial plane backward in the position of the emitter with negative velocities ${\theta'}_{o}$ have greater chance of escaping the black hole than the ones that have positive velocities ${\theta'}_{o}$. In analogy to the Newtonian approximation of the multiple moments it appears that, the gravitational field of the upper plane is stronger than the lower one. This ``asymmetry'' of the gravitational field creates the illusion that the black hole is separated in two parts, as if the two masses on the axis of the black hole on the upper and lower plane which are creating the distortion field are ripping apart the black hole into two pieces. Note that the sources of distortion are not included in  the solution;  the spacetime region under consideration is vacuum. As $c_3$ gets smaller the primary shadow moves to the centre and the secondary image  gets smaller and disappears for $\left|c_3\right| = \left|c_{crit}\right| =\left|1/998\right|$.

While our metric represents a vacuum space-time, there must be (more distant) matter sources causing distortion of the black hole. In this work, we consider the distorted black hole as a local solution, which is valid only in a certain region interior to the sources, far away from the sources, and exterior to the horizon. A global solution can be constructed if the solution is extended to an asymptotically flat solution, beyond the sources, by some sewing technique. Provided that the distorting sources are located in some finite-sized region we can extend the solution, by cutting the spacetime manifold in the region where the metric is valid and attaching to it another spacetime manifold where the solution is not vacuum anymore, and  extending it to yet a more distant vacuum asymptotically flat region. 

We close this section with some comments for the case $r_e>r_o$. Extra photons might be captured in this case, leading to extra structure in the original shapes, since as $r_e$ increases the criteria for escape from the black hole becomes  increasingly strict. Therefore, more points may belong to the shadow, but no point belonging to the shadow can disappear. In particular, the points belonging to the secondary shadow remain black.  However, we have made a limited investigation for the $r_e>r_o$ case and did not find any evidence of merging of the primary shadow to the eyebrow-like structure. However, we cannot increase $r_e$ much, since the distortion fields grow as we go to larger radius and the criteria $f=g_{tt(d)}/g_{tt}<10$ would be gradually violated at some radius, depending on the value of the multiple moment $c_3$. 

\section{Summary}

One striking feature associated with the shadow of multi-black holes is an eyebrow-like structure.  This structure has also been observed for the shadows of two merging black holes, and has generally been regarded as a unique feature of multi black hole systems. In the case of two black holes each eyebrow-like shadow corresponds to the rays getting bent by one black hole that eventually get captured by the other black hole. Geodesics traced from one black hole must be deflected around the other black hole to generate an eyebrow. In particular, one observes two primary shadows corresponding to the two black holes and  at least two eyebrow-like shadows. 

We have shown that, such eyebrow-like structure can also appear for a single distorted   black hole. For the specific case of octupole distortions (with $c_3=-c_1$), an observer sees two shadows or two images for a the single black hole: a larger primary image and a secondary smaller one. 

We considered the case of the octupole distortion of a  black hole with $c_3=-c_1$, concentrating on the situation where $g_{ttd}/g_{tt} < 10$ implying the distorted spacetime is not much different from the undistorted one.  A null ray traced back from the point $(r_{o},\theta_{o})$, with initial parameters $(p, q)$, was taken to be part of the local shadow if it got absorbed by the black hole, or in other words  approached the horizon sufficiently closely. However if it  reached a radius $r_{e}$ (after propagating in the space-time) it was not considered to be part of  the ``local shadow'' of the black hole. 
 
 The parameters $(p, q)$ are related to the initial velocity ${\theta'}_{o}$, integral of motion $l_z$, and the position of the observer. As we saw from the results, for an observer in the equatorial plane the shapes of the black hole shadows are related with a transformation $c_3 \rightarrow -c_3$ and $p \rightarrow -p$. 

The secondary image, or eyebrow, of the black hole shadow is a smaller,  almost mirror reflected shape of the primary shadow image, with respect to the horizontal axis. For $c_3 >0$, more photons with ${\theta'}_{o}>0$ are getting absorbed from the black hole than photons with ${\theta'}_{o}<0$.  It appears that, as in the Newtonian case, the gravitational field in the upper plane is stronger than the lower one. The fact that in the $(p,q)$ plane the primary shadow appears in the lower plane for $c_3 >0$ has to do only with the choice of sign between the relation of ${\theta'}_{o}$ and $p$. 

Last but not least, the creation of two images gives the illusion of   separation of the black hole in two parts. However, as $c_3$ decreases the primary shadow moves to the centre, as we can see  from the graphs for the trajectories of a photon in the distorted spacetime.  The secondary image gets smaller and disappears for $\left|c_3\right| = \left|c_{crit}\right| =\left|1/998\right|$. 

To fully resolve every shadow it would require infinitely small pixels, something beyond the scope of our ability. Consequently, we cannot exclude the existence of other, even smaller shadows, which would correspond to multiple images of the shadow rather than double images. Such features have been observed in the simulations of merging black holes \cite{merge}.  t would be interesting to investigate whether they are also present for distorted black holes.

\section*{Acknowledgment}
The authors are grateful to the Natural Sciences and Engineering Research Council of Canada for financial support. The authors would like to acknowledge Don N. Page for  useful comments. We also thank Andy Bohn for answering some of our questions regarding their paper \cite{merge}. S.A. and C.T. would like to thank Perimeter Institute and University of Waterloo for their hospitality in Waterloo, Ontario, Canada.

\appendix

\section{Newtonian multiple moments}

In this appendix, we calculate what are the Newtonian multiple moments associated with the potential of external sources $U$. The potential (\ref{Upot}) can be written in the terms of two Legendre polynomials,
\be
U(\theta , r)=\sum_{n=0}^{\infty}q_{n}P_{n}(\cos\theta)P_{n}(2r-1) \n{secformofU} \, ,
\ee the relations between the coefficient of the (\ref{Upot}) and the (\ref{secformofU}) when $c_1=-c_3$ are,
\be\label{cns}
c_0= q_0-\frac{q_2}{2} ~~,~~ c_2=\frac{3q_2}{2} ~~,~~ c_1=-c_3=-\frac{5q_3}{2}
\ee 
For more details and derivation of these relations see \cite{AnSh}. In cgs units, the argument of the second Legendre polynomial can be written as,
\be
2r-1=\frac{\rho-M\lambda G}{M\lambda G} ~~~~ , ~~~ \lambda=\frac{1}{c^2} \, ,
\ee where $c$ is the speed of light and $G$ is the gravitational constant. Therefore,
\be
U(\theta , \rho)=\sum_{n=0}^{\infty}q_{n}P_{n}(\cos\theta)P_{n}\left(\frac{\rho-M\lambda G}{M\lambda G}\right) \, .
\ee We can expand the Legendre polynomials $P_{n}\left(\frac{\rho-M\lambda G}{M\lambda G}\right)$ as following:
\be
P_{n}\left(\frac{\rho-M\lambda G}{M\lambda G}\right)=\sum_{l=0}^{n} A_{n \, l} \left(\frac{\rho-M\lambda G}{M\lambda G}\right)^l ,
\ee where 
\be
 A_{n \, l}\equiv 2^n {n\choose l}{\frac{n+l-1}{2}\choose n}. 
 \ee
The Newtonian moments can be calculated by using the coordinate-invariant Ehlers definition \cite{Ehlers}, (for the calculation of  Newtonian moments associated to the exterior multiple moments see \cite{MM4}), 
\be
\Phi=\lim_{\lambda\rightarrow 0}\frac{1}{\lambda}U(\theta , \rho)=\lim_{\lambda\rightarrow 0}\frac{1}{\lambda}\sum_{n=0}^{\infty}q_{n}P_{n}(\cos\theta)\sum_{l=0}^{n} A_{n \, l}\left(\frac{\rho-M\lambda G}{M\lambda G}\right)^l \, .
\ee Based on dimensional analysis we introduce, $\tilde{q}_{n}={q}_{n}\left(\frac{1}{\lambda G}\right)^{n+1}\frac{1}{M^n}$. Thus, we have the following expression for $n=l$,
\be
\Phi=G\sum_{n=0}^{\infty}\tilde{q}_{n}A_{n \, n} \, P_{n}(\cos\theta)\rho^n \, ,\n{A8}
\ee We have $A_{n \, l}$ in factorial form,
\be
A_{n \, l}=2^n\frac{n!}{l!\left(n-l\right)!}\frac{\left(n+l-1\right)!}{l!\left(n-1\right)!} \, ,
\ee for $n=l$,
\be
A_{n \, n}=2^n\frac{\left(2n-1\right)!}{n!\left(n-1\right)!} \, ,
\ee for $n=0$ we get, $A_{n \, n}=1$. 

 The expansion of the Newtonian potential in terms of Legendre polynomials inside a gravitational mass distribution, when radius $\rho$ of the observation point is less than the radius of the mass source, is written as,
\be
U_N=G\sum_{n=0}^{\infty}\tilde{c}_{n} P_{n}(\cos\theta)\rho^n \, .\n{A11}
\ee Here, $\tilde{c}_{n}$'s are the interior Newtonian multiple moments and $\theta$ is the observation angle. Comparing (\ref{A8}) and (\ref{A11}), we conclude that,
\be
\tilde{c}_{n}=\tilde{q}_{n}A_{n \, n} = A_{n \, n}\, {q}_{n}\left(\frac{1}{\lambda G}\right)^{n+1}\frac{1}{M^n} \, ,
\ee in $G=c=1$ units we have,
\be
\tilde{c}_{n}=A_{n \, n}\, {q}_{n}\frac{1}{M^n} \, .
\ee
Let us now consider the example of the Newtonian potential of two masses $m_1$ and $m_2$, located at distances $\rho_1$ on the upper plane and $\rho_2$ on the lower one, from the centre on the axis, respectively, and a thin ring of mass $m$
and radius $a$ on the equator. The potential of this configuration at a point $(\rho,\theta )$ is 
\be
U_{N}=-\frac{m_1}{B_1}-\frac{m_2}{B_2}-\frac{2mK(s)}{\pi B_3} \, ,
\ee 
where,
\ba
B_1&=&\left[{\rho}^{2}+{\rho_1}^2-2\, \rho\,\rho_1\,\cos\theta\right]^{\frac{1}{2}} \, ,~~~
B_2=\left[{\rho}^{2}+{\rho_2}^2+2\, \rho\,\rho_2\,\cos\theta\right]^{\frac{1}{2}} \, ,\nonumber\\
B_3&=&\left[{\rho}^{2}+{a}^2+2\, a\, \rho\,\cos\theta\right]^{\frac{1}{2}}\, ,~~~
s=\frac{2\sqrt{a \rho \,\sin\theta}}{B_3}\, ,
\ea  and $K(s)$ is the complete elliptic integral of the first kind, $K(0)=\pi/2$.
Consider an approximation of the potential under the assumption that, the sources are remote and the potential is calculated in the interior of the sources, (i.e., $\rho_1$ and $\rho_2$, and $a$ are much greater than $\rho$). Then  
\be\label{ctns}
U_{N}=\tilde{c}_0+\tilde{c}_1\, \rho \, P_1(\cos\theta)+\tilde{c}_2\, {\rho}^2 \, P_2(\cos\theta)+\tilde{c}_3\, {\rho}^3 \, P_3(\cos\theta')+ ... \, .
\ee 
where 
\ba
\tilde{c}_0&=&-\frac{m}{a}-\frac{m_1}{\rho_1}-\frac{m_2}{\rho_2} \, , ~~~
\tilde{c}_1=-\frac{m_1}{{\rho_1}^2}+\frac{m_2}{{\rho_2}^2} \, , \nonumber\\
\tilde{c}_2&=&\frac{m}{2a^3}-\frac{m_1}{{\rho_1}^3}-\frac{m_2}{{\rho_2}^3} \, ,~~~
\tilde{c}_3=-\frac{m_1}{{\rho_1}^4}+\frac{m_2}{{\rho_2}^4} \, .
\ea are the Newtonian multiple moments. In the Newtonian case with proper choice $m_1$, $m_2$, $m$, $\rho_1$, $\rho_2$ and $a$ we can have $\tilde{c}_2=0$ and $\tilde{c}_1=-\tilde{c}_3$.  Using   the analogy between $U_{N}$ in the Newtonian case and $U$, we can interpret the $c_n$ multipole moments in (\ref{cns}) in terms of the $\tilde{c}_n$ moments in (\ref{ctns}).


\end{document}